\documentclass[12pt,a4paper]{article}
\usepackage{jheppub}
\usepackage{dsfont}
\usepackage{amssymb, amsmath}
\usepackage{epsfig}
\usepackage{epstopdf}
\usepackage{latexsym}
\usepackage{graphicx}
\usepackage{subfigure}
\usepackage{booktabs}
\usepackage{bbm}
\pdfoutput=1
\makeatletter
\def\@fpheader{\relax}
\makeatother

\def\T{{\cal T}}

\def\F{{\bar {F}}}
\def\M{{\cal M}}
\def\N{{\cal N}}

\def\S{{\cal S}}
\def\A{{\cal A}}

\def\L{{\cal L}}
\def\O{{\cal O}}
\def\J{{\cal J}}
\def\cF{{\cal F}}
\newcommand{\be}{\begin{equation}}
\newcommand{\ee}{\end{equation}}
\newcommand{\bea}{\begin{eqnarray}}
\newcommand{\eea}{\end{eqnarray}}
\newcommand{\bear}{\begin{eqnarray}}
\newcommand{\eear}{\end{eqnarray}}
\newcommand{\beas}{\begin{eqnarray*}}

\newcommand{\eeas}{\end{eqnarray*}}
\newcommand{\ba}{\begin{array}}
\newcommand{\ea}{\end{array}}

\title{Flavour Fields in Steady State: Stress Tensor and Free Energy}
\author{Avik Banerjee$^{a}$,~Arnab Kundu$^{a}$,~ Sandipan Kundu$^{b}$}
\affiliation{$^a$Theory Division, Saha Institute of Nuclear Physics, 1/AF Bidhannagar, Kolkata- 700064, India.}
\affiliation{$^b$Department of Physics, Cornell University, Ithaca, New York, 14853, USA.}
\emailAdd{avik.banerjee [at] saha.ac.in}
\emailAdd{arnab.kundu [at] saha.ac.in}
\emailAdd{kundu [at] cornell.edu}

\abstract{The dynamics of a probe brane in a given gravitational background is governed by the Dirac-Born-Infeld action. The corresponding open string metric arises naturally in studying the fluctuations on the probe. In Gauge-String duality, it is known that in the presence of a constant electric field on the worldvolume of the probe, the open string metric acquires an event horizon and therefore the fluctuation modes on the probe experience an effective temperature. In this article, we bring together various properties of such a system to a formal definition and a subsequent narration of the effective thermodynamics and the stress tensor of the corresponding flavour fields, also including a non-vanishing chemical potential. In doing so, we point out a potentially infinitely-degenerate scheme-dependence of regularizing the free energy, which nevertheless yields a universal contribution in certain cases. This universal piece appears as the coefficient of a log-divergence in free energy when a space-filling probe brane is embedded in AdS$_{d+1}$-background, for $d=2, 4$, and is related to conformal anomaly. For the special case of $d=2$, the universal factor has a striking resemblance to the well-known heat current formula in $(1+1)$-dimensional conformal field theory in steady-state, which endows a plausible physical interpretation to it. Interestingly, we observe a vanishing conformal anomaly in $d=6$.}

\begin{document}

\maketitle
\flushbottom

\section{Introduction \& Discussion}

Thermodynamics and its underlying microscopic description, equilibrium statistical mechanics, have taught us a wealth of information for various physical phenomena across a remarkably large range of energy-scales. These include the existence of various phases of matter, their intricate interplay, corresponding stability and as far as dissipation of diverse {\it small} fluctuation modes, that is connected to an equilibrium thermodynamical description {\it via} a linear perturbation theory. It remains unclear how to explore regimes which are not perturbatively connected to equilibrium, specially due to the lack of an understanding of the underlying principles, if there exists any. We will generically call this a {\it far-from-equilibrium} situation.

On the other hand, many physical systems and situations are inherently dynamical and require a good control over the far-from-equilibrium physics. In classical physics, fluctuation relations\cite{GallavottinCohen} and large deviation techniques\cite{Kurchan, LebowitznSpohn} have become powerful methods in addressing such issues. Quantum mechanically, however, a systematic treatment is lacking till date.

Given a quantum field theory, one may address real-time dynamical issues using Schwinger-Keldysh formalism. In practice though, this is generally a hard problem. Furthermore, this formalism is based on summing up Feynman diagrams up to a certain order, hence perturbative. This perturbative treatment relies on the existence of a small dimensionless coupling in the theory. Ideally though, one would like to have a framework in which dynamical issues can be addressed beyond the perturbation scheme. The strongly coupled physics of the Relativistic Heavy Ion Collider (RHIC) and the Large Hadron Collider (LHC), and numerous strongly coupled condensed matter systems provide us with physical scenarios in which such a framework is essential. Thus a set of toy-models where similar issues may be explored, will be a welcome addition.

Interestingly, from a theoretical perspective, an incredible amount of knowledge is generally gathered for conformal field theories (CFT). The latter is physically relevant near phase transitions, in describing critical phenomena. These, nonetheless, provide us with a powerful ground -- aided by the presence of conformal symmetry -- where various physical issues can be explored, often understood in details. Hence, CFTs are natural candidates where {\it far-from-equilibrium} dynamics can begin to be probed. For our current purposes, we will restrict ourselves to a certain non-equilibrium steady-state (NESS) situation.

Typically, a static or purely equilibrium system is described by a set of time-independent macroscopic variables, both intensive and extensive. However, these macroscopic variables do not have the notion of a {\it velocity} or a {\it flow} inherently associated with them. Steady-state systems, on the other hand, require time-independent variables which intrinsically carry intimation of a {\it flow}, {\it e.g.}~the ones associated with a charge-current or a heat-current. In our case, we will consider a non-vanishing charge-current flow, which will have an effective thermodynamic description characterized by an intensive variable (specifically, an effective temperature), and the conjugate extensive one (specifically, the entropy)\cite{Kundu:2013eba}.\footnote{Note that, in \cite{Kundu:2015qda} it was mentioned that such an effective thermodynamic description may not have a well-defined notion of entropy. A better way of stating the fact is: one can define an entropy, as in \cite{Kundu:2013eba}, but in the gravitational description this entropy does not correspond to the area of an event-horizon; unlike usual cases.}

Before proceeding further, let us note that in \cite{BD1, BD2} the existence of a steady-state configuration has been established rigorously for $(1+1)$-dimensional conformal field theory. Furthermore, the existence of a universality has been argued.\footnote{Similar universality relations have already been conjectured in \cite{Chang:2013gba}, for a $d$-dimensional CFT for $d \ge 3$. This has further been validated in the holographic context in \cite{Amado:2015uza}.} We, however, will not strictly work with a conformal field theory; instead we will introduce a {\it probe degree of freedom} propagating in the background of a conformal system. This is more in keeping with studies in defect conformal field theory. The non-vanishing charge-current will be induced in this defect sector, that subsequently sets up the steady state configuration. Let us now be more specific.

We have remarked on the necessity of an in principle non-perturbative framework, where strong coupling dynamics can be examined. To this end, we will use Gauge-String duality, that relates classical gravity with strongly coupled quantum systems. More specifically, in this article, we will use the best-understood example of this duality, the AdS/CFT correspondence\cite{Maldacena:1997re}, wherein a probe brane will subsequently be introduced in the classical geometry of asymptotically Anti de-Sitter (AdS) space. This corresponds to adding defect degrees of freedom in the dual conformal field theory. But our discussion is certainly not limited to only asymptotically AdS-spaces, or, conformal field theories.

Typically, the picture is as follows: We take $N_c$ number of {\it e.g.}~D$3$-branes, with $N_c \to \infty$, stacked together and placed on the tip of a cone (Calabi-Yau $3$-fold) whose base can simply be an Einstein manifold or a Sasaki-Einstein manifold, if we want to preserve some supersymmetry.\footnote{This can be generalized for D$p$-branes, in which case the base can either be simply Einstein manifolds, or Sasaki-Einstein manifolds or nearly K\"{a}hler manifolds. The latter choices preserve some amount of supersymmetry.} The near-horizon limit of this configuration yields a geometry of the form AdS$_5 \times \M^5$, where $\M^5$ is a compact $5$-dimensional manifold, such as the $L^{p,q,r}$-family of Sasaki-Einstein manifolds. The corresponding dual gauge theories are the so called {\it quiver gauge theories} with a $\prod_{i=1}^{p+q} SU(N_c)$-gauge group. With the same supergravity field content, it is easy to introduce a black hole in the geometry and therefore study the finite temperature version of the corresponding CFT.

In this generically $\N = 2$ supersymmetric CFT background we can introduce probe flavour degrees of freedom that transform under the fundamental representation of the gauge group. This can be done following the approach in \cite{Karch:2002sh}: We introduce $N_f$ probe D$7$-branes, in the limit $N_f \ll N_c$, that span the ${\mathbb R}^{1,3}$-directions inside the AdS and wrap an $\M^3 \subset M^5$. The dynamics of this probe sector is governed by the Dirac-Born-Infeld (DBI) action, along with an Wess-Zumino term. The steady-state is arranged by exciting a constant electric field on the worldvolume of the probe, which drives a non-zero flavour-current.

One may consider a time-dependent situation beyond the probe limit, in which there is an evolution between an initial and a final state. In the dual gravitational picture, the stress-energy tensor of the probe branes will back-react on the AdS geometry, and will presumably back-react in a time-dependent manner, to produce an interpolating dynamical geometry. We will need the stress-energy tensor of the probe sector to obtain this dynamical geometry.

Towards that end, in this article, we explore the stress-energy tensor of the probe sector. Note that, in \cite{Karch:2008uy}, the authors have already explored various aspects of this. Using \cite{Karch:2008uy} as the groundwork, we will rewrite some of their results, specifically for a steady-state scenario. We will observe that the stress-energy tensor of the flavour sector --- that amounts to adding open string degrees of freedom --- is most naturally written involving, but not exclusively, the open string data: the open string metric\cite{Seiberg:1999vs}. In doing so, we will naturally circumvent the IR-divergence encountered in \cite{Karch:2008uy}.

In \cite{Kundu:2013eba, Kundu:2015qda}, it has been argued that an open string equivalence principle gives rise to an effective temperature description in the flavour sector steady-state. It was further argued in \cite{Alam:2012fw} that a well-defined notion of (Helmholtz) free energy can also be defined by appropriately choosing the infrared cut-off, still working within the otherwise standard lore of Euclidean quantum gravity.\footnote{Note that, as elegantly argued in \cite{Karch:2008uy}, this is the only extensive quantity that can be reliably calculated in the probe limit. For completeness, we will review their argument in a later section.} This particular infrared cut-off is crucial to avoid the issues, {\it e.g.}~in \cite{Johnson:2008vna}, with naively identifying the free energy as the on-shell action of the probe-brane action. These observations and suggestions were made emboldened by earlier studies in \cite{Albash:2007bq, Kim:2011qh, Sonner:2012if}, which explicitly demonstrated various qualitative similarities of the flavour dynamics in a background electric field as compared to a thermal background.

In this article we attempt to tailor together a consistent story of the effective thermal description by bringing in various existing pieces of evidence to a formal structure. In writing down the stress-tensor\footnote{We are referring to both the bulk gravitational stress-tensor and the dual field theory one.} of the flavour sector in terms of the open string metric, it becomes apparent why a new and unique infrared cut-off naturally emerges in terms of the so-called open string metric. Subsequently we define the corresponding thermal description by first characterizing the corresponding ensemble in terms of the intensive variables, and then the extensive quantities such as the free energies.

In doing so, we encounter the following non-trivial features: First, it is known that in the presence of a finite charge density (or a chemical potential), the open string metric acquires an ``ergo-plane" and the event horizon generating Killing vector is different from the time-translation generating Killing vector. This alters the regularity condition on the appropriate one-form on the bifurcation surface. As a result, the chemical potential of the flavour sector becomes a function of the adjoint sector temperature, and the applied electric field. Equivalently, the chemical potential depends on two temperatures: the background adjoint-sector one and the effective one. We demonstrate the dependence on the electric field with simple examples.

The second non-trivial feature is related to the regularization of the Euclidean on-shell action of the probe, which is identified as the corresponding free energy. Ordinarily, given a geometry, one considers a constant radial-slice of the metric and subsequently writes down covariant counter-terms on this slice. In the present case, this is subtle since we have two geometries: the background gravity metric, and the open string metric. It is straightforward to check that covariant counter-terms written on a constant radial-slice of both these metrics cancel the UV-divergences, but contribute differently in the IR. Therefore, the regularization process is scheme-dependent: we can use the closed string metric ({\it e.g.}~an AdS-geometry) or the emerging open string metric. In fact, we can ``mix" these two classes of covariant counter-terms in an infinite number of ways; therefore the regularization is not only non-unique, it is infinitely degenerate. We will, nevertheless, use the closed string metric in writing down the counter-terms, just for simplicity.

An interesting situation arises in examples where the background is AdS$_3$ and AdS$_5$: there are log-divergences in the free energy and the trace of the flavour field stress tensor. This, by definition, introduces an arbitrary scale in the regularization process. The origin of this log-divergence is the presence of a conformal anomaly\footnote{We would expect a non-vanishing conformal anomaly for odd-dimensional AdS-geometries. However, an explicit computation shows that no such log-divergence appears in AdS$_7$ which is dual to a six-dimensional CFT. We will not offer a resolution of this in this article.}, which subsequently yields a non-zero trace of the flavour field stress tensor. The coefficient of the log-divergence, however, does not depend on the regularization process and is completely unique. In AdS$_3$-background, which is dual to an $(1+1)$-dimensional CFT, this coefficient is proportional to: $\left( T_{\rm eff}^2 - T^2 \right)$, where $T_{\rm eff}$ is the effective temperature and $T$ is the temperature of the AdS$_3$-BTZ geometry. This is tantalizingly similar to the steady-state heat current result obtained in $(1+1)$-dimensional CFT in \cite{BD1, BD2}, purely from the symmetry of the system. Note that, our set up is different from theirs; nevertheless, it is interesting to note the similarity and conjecture that the universal piece in the free energy ({\it i.e.}~coefficient of the log-divergence) corresponds to a heat-current. In AdS$_5$-BH background, this universal piece takes a more complicated form and is no longer a simple linear combination of two black-body--like radiation.

Given the free energy, we can define the corresponding entropy by taking an (effective) temperature derivative\cite{Kundu:2013eba}. It is further possible to express this result in a completely covariant form involving the open string metric and the background metric. However --- because this thermodynamic entropy involves both the open string metric and the background metric --- it is straightforward to check that the conjugate entropy is not given by the area of the open string metric event horizon, and the physical interpretation of the ``entropic quantity" ({\it i.e.}~as defined in \cite{Kundu:2015qda} in terms of the area) remains elusive. Thus, we will not dwell more on entropy and leave it for future work.

This article is divided in the following parts: In the next section we set up the framework and the basic ingredients. We discuss the appearance of open string metric horizon in bulk stress-tensor, field theory stress-tensor and the free energy in section 3. We also define the thermodynamic ensemble and the corresponding intensive variables here. Then we demonstrate various features that we have summarized above, with examples in section 4. Section 5 is devoted to an at-length discussion on the special case of AdS$_3$-background. Finally, we end with a review of the toy model of \cite{Karch:2008uy} in section 6 for clarity and completeness.

\section{Steady state system and open string metric}

In this section we will set the stage for subsequent discourse. To begin with, we will keep the discussion general, ready to be opted for a D$q$-brane probing a generic supergravity background. Later on we will focus on asymptotically AdS$_{d+1}$-spaces as the prototypical supergravity background which has $(d+1)$ non-compact directions. Much of what we discuss here is already contained in \cite{Kundu:2013eba, Kim:2011qh}. For the sake of completeness, we will include some explicit formulae, which have already appeared in {\it e.g.}~\cite{Kim:2011qh}.

Let us start from a $10$-dimensional closed-string geometry, in Einstein frame\footnote{The metric components in string frame and Einstein frame are related by: $G_{\mu\nu}^{({\rm E})} = e^{-\phi/2} G_{\mu\nu}^{({\rm string})}$, where the superscripts are self-explanatory. For simplicity, however, we will drop the superscript $(\rm E)$ but will always work in the Einstein frame.}:
\begin{eqnarray}
ds^2 & = & G_{tt} dt^2 + G_{xx} dx^2 + G_{yy} d\vec{y}_{d-2}^2 + G_{uu} du^2 +  G_{c}d\Omega_{9-d}^2 \ , \label{csmet1} \\
\phi & = & \phi(u) \ , \label{csmet2}
\end{eqnarray}
where we are suppressing all other geometric data, {\it e.g.}~various non-vanishing fluxes that may be present in the background. The various metric functions, collectively denoted by $G$, are functions only of the radial coordinate $u$. The putative dual $d$-dimensional field theory has an ${\mathbb R}_t \times {\mathbb R}_x \times SO(d-2)$ symmetry, which is enhanced to ${\mathbb R}_t \times SO(d-1)$ if $G_{xx} = G_{yy}$.

Suppose now, that in the closed-string background (\ref{csmet1})-(\ref{csmet2}), we introduce a probe ($N_f$ number of) D$q$-brane that, including $x$, wraps $m$ of the field theory spatial directions, the time direction $t$, along with the radial direction $u$ and a $(q-m-1)$ submanifold inside the $(9-d)$ compact manifold. Clearly, this is possible as long as $\left( d + q -m \right) \le 10$. The action for the probe brane is given by
\begin{eqnarray}
&& S_{{\rm D}q} = - N_f \tau_q \int d^{q+1} \xi \ e^{\gamma \phi} \sqrt{- {\rm det} \left[ P[G + B] + \left( 2 \pi \alpha' \right) e^{-\phi/2} F\right]} + S_{\rm WZ} \ , \\
&& {\rm with} \quad \gamma = \frac{q-3}{4} \ ,
\end{eqnarray}
where $\tau_q$ denotes the tension of a single D$q$-brane, $G$ and $B$ denote the metric and the NS-NS $2$-form, $P[\ldots]$ denotes the pull-back, $\alpha'$ denotes the string tension and $F$ denotes the $U(1)$-gauge field living on the worldvolume of the D$q$-brane. We will not be explicit about the Wess-Zumino terms, denoted by $S_{\rm WZ}$, since they would not affect our discussion. For simplification, we will assume that $B=0$ and the brane embedding is trivial, {\it i.e.}~$P[G] = G$ when evaluated on the probe.\footnote{These are simplifying assumptions. If we have a non-zero $B$, it can be easily packaged into a new anti-symmetric $2$-form: $\Sigma_2 = P[B] + \left( 2 \pi \alpha' \right) e^{-\phi/2} F$. Everything that follows can subsequently be written in terms of $\Sigma_2$. The assumption of a trivial embedding, on the other hand, is more than a notational convenience. We will later comment on (some of) the physical results that depend on this assumption.}

In what follows we will focus only on the DBI-part of the action. Let us now specifically consider the case where we take the following ansatz for the $U(1)$ gauge field on the probe:
\begin{eqnarray}
F = - E dt \wedge dx + a_x'(u) du \wedge dx + a_t'(u) du \wedge dt  \ , \label{Fansatz}
\end{eqnarray}
where $E$ is a constant field, $a_x(u)$ and $a_t(u)$ are to be determined from the corresponding equations of motion.\footnote{For the critical reader, we remark that the ansatz in (\ref{Fansatz}) is compatible with the completely covariant equation of motion for $F$ that is obtained from the DBI. Evidently we are implicitly ignoring the Wess-Zumino term in all these discussions. Typically, a non-vanishing Wess-Zumino term can induce more gauge-components on the probe worldvolume. We will not address this possibility here.} Thus the DBI-part of the action yields
\begin{eqnarray}
S_{\rm DBI} = - N_f \tau_q \int d^{q+1} \xi \ e^{\left(\gamma-1/2\right)\phi} && \left( - G_{uu} \left( e^{\phi} G_{tt} G_{xx} + E^2 \right) - G_{tt} a_x'(u)^2 - G_{xx} a_t'(u)^2 \right)^{1/2}  \nonumber \\
&& \left( G_{yy}^{(m-1)} G_c^{(q-m-1)} \right)^{1/2} \ .
\end{eqnarray}
The equations of motion for $a_x(u)$ and $a_t(u)$ are given by
\begin{eqnarray}
e^{\left(\gamma-1/2\right)\phi} \frac{\left( G_{yy}^{(m-1)} G_c^{(q-m-1)} \right)^{1/2}G_{tt} a_x'(u)}{\sqrt{ - G_{uu} \left( e^{\phi} G_{tt} G_{xx} + E^2 \right) - G_{tt} a_x'(u)^2 - G_{xx} a_t'(u)^2 }} & = & j \ , \label{eomax} \\
e^{\left(\gamma-1/2\right)\phi} \frac{ \left( G_{yy}^{(m-1)} G_c^{(q-m-1)} \right)^{1/2} G_{xx} a_t'(u)}{\sqrt{ - G_{uu} \left( e^{\phi} G_{tt} G_{xx} + E^2 \right) - G_{tt} a_x'(u)^2 - G_{xx} a_t'(u)^2 }} & = & d_0 \ . \label{eomat}
\end{eqnarray}
Here, eventually upon implementing the AdS/CFT dictionary, the constant $j$ and $d_0$ are related to the flavour current and flavour charge density, respectively. The steady-state physics is completely contained in the applied electric field $E$ and the response current $j$, even in the absence of a charge density, where it is purely governed by Schwinger-type pair production\cite{Karch:2007pd}.

In \cite{Kundu:2013eba, Kundu:2015qda}, it has been argued that the flavour sector, {\it i.e.}~the open string degrees of freedom, is coupled to the open string metric which is given by
\begin{eqnarray}
S_{ab} = G_{ab} - \left( \bar{F} G^{-1} \bar{F} \right)_{ab} \ , \quad \bar{F} = \left( 2 \pi \alpha' \right) e^{-\phi/2} F \ .
\end{eqnarray}
Here $\bar{F}$ denotes the worldvolume $U(1)$-gauge field and $a,b, \ldots$ are used to describe worldvolume directions. In our subsequent discussions, we will set $2 \pi \alpha' =1$, and thus choose a corresponding energy-unit in which we measure all bulk dimensionful quantities. Turned into a dual field theory statement, this amounts to measuring dimensionful quantities in units of the 't Hooft coupling which is defined in terms of the string coupling $g_s$ and the inverse string tension $\alpha'$. Since we are keeping the discussion as general as possible, we will not make this choice more precise.

Evaluated on the background in (\ref{Fansatz}), the open string line element is 
\begin{eqnarray}
ds_{\rm osm}^2 & = & \left( G_{tt} G_{xx} + e^{-\phi} E^2 \right) \left[ \frac{d\tau^2}{G_{xx}} + \frac{dX^2}{G_{tt}} \right] + \frac{e^{-\phi}}{G_{uu}} \left( a_t' d\tau + a_x' dX \right)^2  \nonumber\\
& + & G_{yy} \sum_{i=1}^{m-1}dy_i^2 + du^2 \left(G_{uu} + e^{-\phi} \frac{a_t'^2 G_{xx} + a_x'^2 G_{tt}}{G_{tt} G_{xx} + e^{-\phi} E^2 } \right) + G_c d\Omega_{q-m-1}^2 \ , \label{osmds} \\
dt & = &  d\tau +  \frac{E a_x'}{e^{\phi} G_{tt} G_{xx} +E^2} du \ , \label{tau2t} \\
dx & = & dX -  \frac{E a_t'}{e^{\phi}G_{tt} G_{xx} +E^2} du \ , \label{Xtox}
\end{eqnarray}
where
\begin{eqnarray}
S_{tt} & = & G_{tt} + e^{-\phi} \left[ \frac{E^2}{G_{xx}} + \frac{a_t'^2}{G_{uu}} \right] \ , \quad S_{xx} = G_{xx} + e^{-\phi} \left[\frac{E^2}{G_{tt}} + \frac{a_x'^2}{G_{uu}} \right] \ , \\
S_{uu} & = & G_{uu} + e^{-\phi} \left[\frac{a_t'^2}{G_{tt}} + \frac{a_x'^2}{G_{xx}} \right] \ , \quad S_{tx} = e^{-\phi} \frac{a_t' a_x'}{G_{uu}} \ , \quad S_{tu} = - e^{-\phi} \frac{E a_x'}{G_{xx}} \ , \\
S_{xu} & = & e^{-\phi} \frac{E a_t'}{G_{tt}} \ , \quad S_{yy} = G_{yy} \ , \quad S_c = G_c \ . 
\end{eqnarray}
This is a generalization of the previously obtained formulae for a D$7$-brane probing a D$3$-brane geometry, in \cite{Kim:2011qh}. The event-horizon is located where the $uu$-component of the metric in (\ref{osmds}) diverges, and is given by
\begin{eqnarray}
e^{\phi} \ G_{tt} G_{xx} +E^2 = 0 \ . \label{osmeve}
\end{eqnarray}
There is a Killing horizon corresponding to the Killing vector $\xi = \xi^\mu\partial_\mu =  \partial/ \partial \tau$, which is given by
\begin{eqnarray}
\frac{G_{tt} G_{xx} + e^{-\phi} E^2}{G_{xx}} + e^{-\phi} \frac{a_t'^2}{G_{uu}} = 0 \ .
\end{eqnarray}
Evaluated on-shell, {\it i.e.}~substituting the equation of motion for $a_t'$, this yields
\begin{eqnarray}
e^{\left(2\gamma -1\right) \phi} G_c^{q-m-1}  G_{yy}^{m-1} G_{tt} + j^2 = 0 \ , \label{osmerg}
\end{eqnarray}
which gives the ergoplane\cite{Kim:2011qh}. Also, the constant $j$ is obtained from the algebraic equation
\begin{eqnarray}
\left. e^{\phi} \left( d_0^2 G_{tt} + j^2 G_{xx} \right)\right|_{u_{*}} + \left. e^{2\gamma\phi} G_c^{q-m-1} G_{yy}^{m-1} G_{xx} G_{tt} \right|_{u_{*}} = 0 \ , \label{osmj}
\end{eqnarray}
where $u_*$ solves open string horizon condition in (\ref{osmeve}).\footnote{It turns out, that imposing a constraint that the osm event horizon is not a Killing horizon with respect to $\xi$ determines the constant $j$. Alternatively, $j$ can also be determined by requiring that the probe brane extends beyond the open string metric horizon.}

\section{Open string metric and stress tensor}

In this section, we will review the method outlined in \cite{Karch:2008uy}, explicitly write down the stress-energy tensor of the flavour sector that eventually back-reacts on the background, explicitly involving the open string metric. This will naturally endorse the choice of the osm event horizon as the natural IR cut-off, proposed in \cite{Alam:2012fw}.

\subsection{Bulk Stress Tensor}

Let us start with the DBI action
\begin{equation}
S_{\rm DBI} = - N_f \tau_q \int d^{q+1} \xi~ e^{\gamma \phi} \left( - \det \left[ G + 2\pi \alpha' e^{-\phi / 2}F \right]\right)^{1/2}
\end{equation}
that, when back-reaction is taken into account, contributes to the right hand side of Einstein's equations. We will rewrite the bulk stress tensor $\T_{MN}$ in terms of the open string data, consisting of the open string metric and its anti-symmetric counter-part. We will assume that $N_f$ $q$-branes have trivial embedding profiles. Furthermore, indices $a,b$ denote coordinates on the worldvolume of $q$-branes, whereas $M,N=0,1,...,9$ are coordinates of the full 10-dimensional bulk.

Now, the bulk stress-energy tensor is\footnote{Note that, if we were working in the string frame the Einstein-Hilbert term would be multiplied by a factor of $e^{-2 \phi}$. This would contribute a factor of  $e^{2\phi}$ in the definition of the probe stress-tensor. Further note that, $\T_{MN} = - \frac{\kappa_{10}^2}{\sqrt{-\det G_{10}}}\left(\frac{\delta S_{\rm DBI}}{\delta G^{MN}} + \frac{\delta S_{\rm DBI}}{\delta G^{NM}}\right)$, {\it i.e.}~with an overall -ve sign as compared to the one with all indices raised. This sign difference comes, essentially from the relation (\ref{MMinvrel}), that translates to $\delta G^{AB} = - G^{AC} \delta G_{CD} G^{DB}$ for the metric.}
\be
\T^{MN} =  \frac{\kappa_{10}^2}{\sqrt{-\det G_{10}}} \left(\frac{\delta S_{\rm DBI}}{\delta G_{MN}}+\frac{\delta S_{\rm DBI}}{\delta G_{NM}}\right) \ . \label{flavourst}
\ee
Here $\det G_{10}$ denotes the determinant of the $10$-dimensional metric, whereas $\det G$ is reserved to denote the determinant of the induced metric on D$q$-branes. 
Thus, as detailed in appendix A, we get
\begin{eqnarray}
\T^{MN} =  - \left(\kappa_{10}^2 N_f \tau_q\right) e^{\gamma \phi} \left( \frac{\sqrt{\det\S\det G}}{- \det G_{10}}\right)^{1/2}\S^{ab} \delta_a^M \delta_b^N \delta^{9-q} \left(\vec{c}-\vec{c}_0\right) \ .
\end{eqnarray}
Here $\vec{c}=\vec{c}_0$ denotes the location of $q$-branes. Thus the bulk stress tensor is, not entirely but up to a conformal factor, expressible in terms of the open string metric\cite{Kim:2011qh}.

\subsection{Boundary Stress Tensor}

It was proposed in \cite{Karch:2008uy} that the boundary stress tensor of the flavour sector (in the dual field theory) should be obtained by taking the variation of the DBI action with respect to the background metric and then integrating each component weighted by the volume factor of the probe D$q$-brane. Evidently, by construction, this quantity is not covariant under $10$-dimensional diffeomorphisms but it is covariant under $(q+1)$-dimensional ones. To set the stage, we review the basic formulae proposed in \cite{Karch:2008uy}.

First, we define
\be
U^{a b } =  \frac{1}{\sqrt{-\det G}} \left(\frac{\delta S_{\rm DBI}}{\delta G_{a b}} + \frac{\delta S_{\rm DBI}}{\delta G_{b a}}\right) \ ,
\ee
where $a, b$ run over the probe worldvolume directions. Clearly, the above definition is made covariant in $(q+1)$-dimensions, but not in the full ten-dimensional embedding space. Also, here $G$ stands for the pull-back of the background on the probe. In what follows, we will only consider the components along the dual ``field-theory directions", which will be denoted by $\mu, \nu$. Thus: $\mu, \nu = 0 , \ldots m$. The stress energy tensor of the boundary theory is given by
\begin{eqnarray}
\langle T^{\mu}_{\nu} \rangle & = & \int_{u_{\rm min}}^{u_{\rm max}} du \,  d^{q-m-1} \vec{c} \ \sqrt{- \det G} \ U^{\mu}_{\nu} \\
& = & - \left( N_f \tau_q \right) \int_{u_{\rm min}}^{u_{\rm max}} du \,  d^{q-m-1} \vec{c} \ e^{\gamma \phi} \left( \frac{\det G}{\det \S} \right)^{1/4} \sqrt{- \det \S} \ \S^{\mu}_{\nu} \\
& = & - \left( N_f \tau_q \Omega_{q-m-1} \right) \int_{u_{\rm min}}^{u_{\rm max}} du \ e^{\gamma \phi} \left( \frac{\det G}{\det \S} \right)^{1/4} \sqrt{- \det \S_0} \ \S^{\mu}_{\nu} \ , \label{tmunuf}
\end{eqnarray}
where we have defined: $\S^{\mu}_\nu = \S^{\mu a} G_{a \nu}$. Here $u_{\rm min}$ and $u_{\rm max}$ are the two limits of the integration that correspond to the UV and the IR of the dual field theory, respectively. Clearly, $u_{\rm min} = 0$,\footnote{Strictly speaking, one needs to set $u_{\rm min} = \epsilon$, introduce counter-terms to cancel the divergences and then take $\epsilon \to 0$ limit. Later we will carry out this procedure in details.} where the conformal boundary is located; the choice of $u_{\rm max}$ is subtle. As argued in \cite{Alam:2012fw}, and we will further endorse this point momentarily, $u_{\rm max} = u_*$, {\it i.e.}~the open string metric event horizon. Furthermore, in (\ref{tmunuf}) we have integrated out the internal directions to obtain a volume factor $\Omega_{q-m-1}$ and $\S_0$ is defined as
\begin{eqnarray}
\int d^{q-m-1} \vec{c} \sqrt{- \det \S}  = \Omega_{q-m-1} \sqrt{- \det \S_0} \ .
\end{eqnarray}

Let us now define the free energy of the system. It was proposed in \cite{Alam:2012fw} that the free energy is given by the Euclidean DBI-action, by sending $t \to - i t_{\rm E}$, up to a boundary term. Omitting this boundary term for now, the Helmholtz free energy density\footnote{In defining the density, we divide by an infinite volume factor: $\int d^m \vec{x} = V_{{\mathbb R}^m}$.} is thus 
\begin{eqnarray}
\cF  =  \left( N_f \tau_q \right) \int_{u_{\rm min}}^{u_{\rm max}} du \ d^{q-m-1} \vec{c} \ e^{\gamma \phi} \sqrt{- \det \M } + {\rm (boundary-term)} 
\end{eqnarray}
Hence, the polarization tensor is defined to be
\begin{eqnarray}
M^{\mu\nu} & = &  - \frac{\partial \cF}{\partial F_{\mu\nu} } = - \frac{\partial \cF}{\partial \left( e^{\phi/2} \bar{F}_{\mu\nu} \right)} \\
& = & - \frac{1}{2} \left( N_f \tau_q \Omega_{q-m-1} \right) \int_{u_{\rm min}}^{u_{\rm max}} du \ e^{\left(\gamma - 1/2 \right)\phi} \left( \frac{\det G}{\det \S} \right)^{1/4} \sqrt{- \det \S_0} \ \A^{\mu\nu} \ , \nonumber\\ \label{Mmunu}
\end{eqnarray}
where $\A^{\mu\nu}$ is the anti-symmetric counter-part of the open string metric and is defined in (\ref{Minv})-(\ref{Aab}). Note that, in the above formula, we could certainly define the polarization tensor using the variation with respect to $\bar{F}$, instead of $F$. However, a fundamental string couples to $F$ and therefore, in the dual field theory, this should be identified with the physical field.

As discussed in \cite{Karch:2008uy}, the stress tensor can be divided into two parts: the fluid part, which encodes the information about dissipation and various transport properties and a polarization part which is non-zero only in the presence of background fields:
\begin{eqnarray}
\langle T^{\mu}_{\nu} \rangle  =  \langle T^{\mu}_{\nu} \rangle_{\rm fluid} + \langle T^{\mu}_{\nu} \rangle_{\rm polarization} \ , \quad  \langle T^{\mu}_{\nu} \rangle_{\rm polarization}  =  M^{\mu\rho} F_{\rho\nu} \ .
\end{eqnarray}
Thus we get
\begin{eqnarray}
\langle T^{\mu}_{\nu} \rangle_{\rm fluid} = - \left( N_f \tau_q \Omega_{q-m-1} \right) \int_{u_{\rm min}}^{u_{\rm max}} du \ e^{\gamma \phi} \left( \frac{\det G}{\det \S} \right)^{1/4} && \sqrt{- \det \S_0} \nonumber\\
&& \left( \S^{\mu}_{\nu} - \frac{1}{2} \A^{\mu\rho} \bar{F}_{\rho\nu} \right) \ . \nonumber\\
\end{eqnarray}
%

\subsection{Open string horizon}

Here we will comment on the lower limit of integration for all the above $u$-integrals. We will argue that the lower limit of integration should be the position of the horizon of the open string metric. As shown in \cite{Karch:2008uy}, there are certain divergence issues when the lower limit of integration is the horizon of the bulk metric. However, all these issues disappear when we move the lower limit to the open string horizon. Since, open-string metric is the quantity that appears in every integral, it is natural to take the open-string horizon as the lower limit. Let us now write down all the above quantities
\begin{align}
&\langle T^{\mu}_{\nu}\rangle = - \left( N_f \tau_q \Omega_{q-m-1} \right) \int_{u_{\rm min}}^{u_{\rm max}} du \ e^{\gamma \phi} \left( \frac{\det G}{\det \S} \right)^{1/4} \sqrt{- \det \S_0} \ \S^{\mu}_{\nu} \ ,\\
& \langle T^{\mu}_{\nu} \rangle_{\rm polarization} = - \frac{1}{2} \left( N_f \tau_q \Omega_{q-m-1} \right) \int_{u_{\rm min}}^{u_{\rm max}} du \ e^{\gamma \phi} \left(\frac{\det G}{\det \S} \right)^{1/4}\sqrt{-\det\S_0} \ \A^{\mu\rho}F_{\rho\nu}\ ,\\
& \langle T^{\mu}_{\nu} \rangle_{\rm fluid} = - \left( N_f \tau_q \Omega_{q-m-1} \right) \int_{u_{\rm min}}^{u_{\rm max}} du \ e^{\gamma \phi} \left(\frac{\det G}{\det \S} \right)^{1/4}\sqrt{-\det\S_0}\left(\S^{\mu}_\nu - \frac{1}{2}\A^{\mu\rho}F_{\rho \nu}\right)\ ,\\
&\cF = \left( N_f \tau_q \Omega_{q-m-1} \right) \int_{u_{\rm min}}^{u_{\rm max}} du \ e^{\gamma \phi}\  \left(\frac{\det G}{\det \S} \right)^{1/4}\sqrt{-\det\S_0} + {\rm (boundary-term)} \ .
\end{align}
It is also important to note that all the above quantities are UV-divergent which can be renormalized by using holographic regularization. We will discuss this later in details. To complete summarizing, let us also note that $u_{\rm min} = \epsilon$ with $\epsilon \to 0$ (corresponding to the UV boundary), and $u_{\rm max}=u_*$ (corresponding to the IR osm horizon). We also emphasize that, following the discussions in \cite{Alam:2012fw}, the identification of $u_{\rm max}$ is also the physically sensible one. It is worth repeating that this choice also removes the IR-divergences discussed in \cite{Karch:2008uy}.

\subsection{Thermodynamics}

Here we begin by reviewing the proposal for the free energy put forward in \cite{Alam:2012fw}, and subsequently generalized in \cite{Kundu:2013eba}. The (Gibbs, and we will justify this momentarily) free energy is identified, as usual, with the Euclideanized DBI-action\footnote{Setting $t \to - i t_{\rm E}$ yields: $S_{\rm DBI} = - N_f \tau_q \int dt \ldots \to + i N_f \tau_q \int dt_{\rm E} \ldots = i S_{\rm DBI}^{({\rm E})}$. We will use this $S_{\rm DBI}^{({\rm E})}$ in our subsequent discussions.} evaluated on-shell. There is, interestingly, a boundary term that originates from the variation of the non-vanishing gauge field\footnote{Note that, after Euclideanization, the time-component of the gauge potential also needs to be analytically continued: $a_t \to - i a_{t_{\rm E}}$. However, all the subsequent thermodynamic formulae will be written in terms of $a_t$, since the on-shell function --- denoted by $a_t$ or $a_{t_{\rm E}}$ --- remains unchanged.} $a_x(u)$:
\begin{eqnarray}
S_{\rm DBI}^{({\rm E})} & = & \frac{N_f \tau_q \Omega_{q-m-1}}{T_{\rm eff}} \int du  \ \L^{({\rm E})} \int d^m\vec{x} \ , \\ 
\left. \delta S_{\rm DBI}^{({\rm E})} \right|_{\rm on-shell} & = & \frac{N_f \tau_q \Omega_{q-m-1} }{T_{\rm eff}} \int d^m\vec{x} \left( \left. \frac{\partial\L^{({\rm E})}}{\partial a_x'} \delta a_x \right|_{\rm boundary}^{\rm horizon} + \left. \frac{\partial\L^{({\rm E})}}{\partial a_t'} \delta a_t \right|_{\rm boundary}^{\rm horizon} \right) \ , \nonumber\\
\end{eqnarray}
where $u_{\rm min}$ and $u_{\rm max}$ correspond to the boundary and the osm horizon, respectively. The variation $\delta a_x$ does not necessarily vanish at the (open string metric) horizon, and thus we need to subtract this boundary term (located at the osm event horizon) to have a well-posed variational problem. This finally yields\cite{Kundu:2013eba}
\begin{eqnarray}
\cF & = & \left.  T_{\rm eff} \ S_{\rm DBI}^{({\rm E})} \right|_{\rm on-shell} + {\rm boundary-terms} \\
& = & \left.  \left( N_f \tau_q \Omega_{q-m-1} \right) \int d^m\vec{x} \left[ \int_{u_{\rm min}}^{u_{\rm max}} due^{\gamma \phi} \left(\frac{\det G}{\det \S} \right)^{1/4}\sqrt{-\det\S_0}\right|_{\rm on-shell}
- \left. \left(j a_x + d_0 a_t \right)\right|_{u_*} \right] \ , \nonumber\\ \label{freeosm} 
\end{eqnarray}
where $j$ and $d_0$ are defined in (\ref{eomax}) and (\ref{eomat}), respectively. Clearly, in the range $u_{\rm max} \le u \le u_{\rm min}$, the $\left( j a_x \right)$, $\left( d_0 a_t \right)$-terms are considered to be usual boundary terms; whereas in the range $u_{\rm min} < u \le \infty$, these act as source terms.\footnote{The reader may notice that, instead of subtracting the boundary term, the gauge field $a_t$ is usually required to satisfy a regularity condition, which takes the form: $a_t ({\rm horizon}) = 0$. Here, we have two candidate horizons: the open string one and the closed string one. Thus, depending on where the regularity is ensured, the corresponding chemical potential will be different. Moreover, because of the presence of an ergo-region, the regularity condition will be non-trivial. We will comment on this momentarily.} Here we have prematurely used the symbol $\cF$ to denote the free energy, which in due course will be identified with the Gibbs' free energy.

We are going to argue that the infrared properties of the system are defined in terms of the open string data, purely. To that end, first note that the following are symmetries of (\ref{osmds}): $\tau \to \tau + {\rm const}$, $X \to X + {\rm const}$, corresponding to the following two Killing vectors: $\xi_{(1)}^\alpha = \{1, 0, 0, \ldots \}$, and $\xi_{(2)}^\alpha = \{ 0, 1, 0, 0, \ldots\}$, wherein the components are ordered as: $\{\tau, X, \ldots\}$ and $\alpha$ denotes indices along the non-compact part of the geometry. Thus, the equation of motion for a probe particle moving in the background of (\ref{osmds}) will result in two conserved quantities. These are:
\begin{eqnarray}
\epsilon & = & - \left( g_{\tau\tau} \frac{d\tau}{d\lambda} + g_{\tau X} \frac{dX}{d\lambda}\right) \ , \\
\ell & = & \left( g_{X\tau} \frac{d\tau}{d\lambda} + g_{X X} \frac{dX}{d\lambda}\right) \ ,
\end{eqnarray}
where $\lambda$ is an affine parameter. Here $\epsilon$ and $\ell$ correspond to energy and momentum, respectively. We have denoted 
\begin{eqnarray}
&& g_{\tau\tau} = \frac{G_{tt} G_{xx} + e^{-\phi} E^2}{G_{xx}} + e^{-\phi} \frac{a_t'^2}{G_{uu}} \ , \\
&& g_{XX} = \frac{G_{tt} G_{xx} + e^{-\phi} E^2}{G_{tt}} + e^{-\phi} \frac{a_x'^2}{G_{uu}} \ , \\
&& g_{\tau X} = e^{-\phi} \frac{a_t' a_x'}{G_{uu}} \ .
\end{eqnarray}

Now we define the ``frame-dragging", characterized by $\Omega(u)$ as follows: Suppose we take $\ell = 0$, this yields
\begin{eqnarray}
\Omega(u)  =   \frac{dX}{d\tau} = - \frac{g_{\tau X}}{g_{XX}}  =   \frac{ d_0 j e^{\phi} G_{yy} G_c^{m+1}} { d_0^2 e^{\phi} G_{yy} G_c^{m+1} + e^{2\gamma \phi} G_{xx} G_{yy}^m G_c^q } \ . \label{dIR1}
\end{eqnarray}
Thus, $\Omega(u_*)$ defines the charge density of the corresponding ensemble, by providing a map of boundary variables $\{d_0, j \}$ in terms of the bulk quantity $u_*$. To define the temperature, let us note that the osm event-horizon is not the Killing horizon corresponding to $\xi_{(1)}$; instead it is a Killing horizon corresponding to\footnote{This result is completely general, without making any reference to the details of the gravity background.} 
\begin{eqnarray}
\xi_{\rm (hori)}^\alpha = \xi_{(1)}^\alpha + \left. \left(-  \frac{g_{\tau X}}{g_{XX}}\right) \right|_{u_*} \xi_{(2)} ^\alpha = \xi_{(1)}^\alpha + \Omega(u_*) \ \xi_{(2)} ^\alpha \ .
\end{eqnarray}
Now it is easy to define the temperature, in terms of $\xi_{\rm hori}^\mu$:
\begin{eqnarray}
T_{\rm eff} = \frac{\kappa}{2\pi} \ , \quad \kappa^2 = \left. - \frac{1}{2} \left( \nabla_\alpha \xi_{{\rm (hori)}_\beta} \right) \left( \nabla^\alpha \xi_{\rm (hori)}^\beta \right) \right|_{u_*} \ . \label{TIR}
\end{eqnarray}

Let us now offer a few comments regarding the triviality of the probe brane embedding. We will work with the well-studied example in which a probe D$7$-brane is introduced in the background of a large number of D$3$-branes. In the dual field theory, it corresponds to introducing ${\cal N}=2$ hypermultiplet in the background of ${\cal N}=4$ super Yang-Mills theory. It is well-known\cite{Mateos:2006nu, Albash:2006ew, Karch:2006bv} that, at finite temperature $T$, depending on the ratio $(m/T)$ --- where $m$ is the mass of the hypermultiplet matter --- there is a first order phase transition from a mesonic state to a plasma state. For $(m/T) \ll 1$, the fundamental matter is in a plasma-phase and for $(m/T) \gg 1$ they exist as bound states.

A similar physics takes place in the presence of an electric field. If one considers the family of probe profiles, such as the ones considered in \cite{Albash:2007bq} --- in which the embeddings are characterized by a mass parameter --- the effective temperature will depend on the profile, and thus in turn, on the mass of the corresponding fundamental field. In \cite{Kim:2011qh}, a general formula was obtained for the effective temperature and one can easily check that increasing the mass parameter results in lowering the effective temperature. Thus, as in the thermally driven phase transition discussed above, we will have bound states for $m/T_{\rm eff} \gg 1 $ and a ``plasma"-phase in the regime $m/T_{\rm eff} \ll 1$. This makes a good qualitative connection with an effective temperature dynamics.\footnote{It should be noted that, as pointed out in \cite{Albash:2007bq}, there exist singular probe embeddings for a certain range of the mass parameter for a non-vanishing electric field. This rather pathological feature is absent in the purely thermal case. Furthermore, the phase transition points, denoted by $m_{\rm cr} / T $ and $ m_{\rm cr} / T_{\rm eff}$, are both $\O(1)$-numbers, which are not equal to each other. So, a quantitative distinction between the thermally driven and the electric-field driven phase transition is immediate. We thank David Mateos for raising this issue.}

Now we want to comment on the chemical potential of the system. To that end, let us return to the regularity of the $1$-form. As we have commented, the osm event horizon in (\ref{osmeve}) is a Killing horizon and therefore it contains the bifurcation surface, with a non-vanishing $\kappa$, where the Killing vector $\partial / \partial Y \equiv \left( \partial / \partial \tau +  \Omega \partial / \partial X \right)$ vanishes\cite{Racz:1992bp}. Hence, any well-defined $1$-form $A_Y$ should vanish at the osm event horizon.\footnote{Clearly, there exists another Killing horizon where $\partial /\partial \tau$ vanishes. It turns out that this Killing horizon also has a non-vanishing surface gravity. However, this ``ergoplane" always lies outside of the osm event horizon and hence we do not need to worry about null geodesic incompleteness at the Killing horizon of $\partial /\partial \tau$.} This will impose $\left. \left( a_t + \Omega a_x \right)\right|_{u_{\rm max} = u_*} =0$.\footnote{This is straightforward to obtain. From definition, we have $dY = d\tau + \Omega^{-1} dX $. Thus $a_Y = a_\tau \frac{\partial\tau}{\partial Y} + a_X \frac{\partial X}{\partial Y} = a_t + \Omega a_x$.} Clearly, this condition does not eliminate the $(d a_t')$-term from (\ref{freeosm}), unlike what is usually encountered. Finally, the chemical potential is defined as:
\begin{eqnarray}
\mu =  \int_{u_{\rm min}}^{u_{\rm max}} F_{ut}  du = \int_{u_{\rm min}}^{u_{\rm max}} a_{t}' du \ . \label{muIR2}
\end{eqnarray}
Here $a_t'$ is the field that appeared in (\ref{Fansatz}). Thus, (\ref{dIR1}) and (\ref{TIR}) characterize the canonical ensemble with a charge density and a temperature, whereas (\ref{muIR2}) defines the chemical potential corresponding to the grand-canonical ensemble.

Before concluding this section let us now comment on the ensemble. To identify the ensemble, let us vary the free energy defined in (\ref{freeosm}), now allowing the variation of the fields at the conformal boundary. 
\begin{eqnarray}
\delta \cF & = & \left. \left( \delta T_{\rm eff} \right) \ S_{\rm DBI}^{({\rm E})} \right|_{\rm on-shell} + \left. T_{\rm eff} \ \delta S_{\rm DBI}^{({\rm E})} \right|_{\rm on-shell}  + \delta \left( {\rm boundary-term}\right) \\
& = &  \left.  \ S_{\rm DBI}^{({\rm E})} \right|_{\rm on-shell} \left( \delta T_{\rm eff} \right)  + \tilde{d}_0 \ \delta \mu + \left. \tilde{j} \ \delta a_x \right|_{u_{\rm min}}  \ , \label{Gibbs}
\end{eqnarray}
where we have set
\begin{eqnarray}
&& \delta \int_{u_{\rm min}}^{u_{\rm max}} a_t' du  = \delta \mu \ , \\
&& \tilde{d}_0 = N_f \tau_q \Omega_{q-m-1} \ d_0  \ , \quad \tilde{j} = N_f \tau_q \Omega_{q-m-1} \ j \ . 
\end{eqnarray}
Note that the variation of the ``boundary term" above cancels the terms coming from the variations at the osm event horizon: in fact, this is the role of the ``boundary terms" in the free energy. It is clear $\cF$ defines the Gibbs' free energy corresponding to the grand-canonical ensemble. The corresponding Helmholtz free energy at vanishing $j$ is defined as:
\begin{eqnarray}
\cF_{\rm H} = \cF - \tilde{d}_0 \ \mu \ ,
\end{eqnarray}
by performing a Legendre transformation to the Gibbs free energy.

\section{Some examples: AdS in $(d+1)$-dimensions}

Let us now focus on some simple examples. Assume that we have a supergravity background of AdS$_{(d+1)} \times \M^{9-d}$, in which $\phi(u)=0$:
\begin{equation}
ds^2 = \frac{L^2}{u^2}\left(du^2+\eta_{\mu\nu}dx^\mu dx^\nu\right) + d\M_{9-d}^2 \ .
\end{equation}
For simplicity, we will set the AdS radius $L=1$. Typically, this choice is equivalent to fixing the Yang-Mills coupling $g_{\rm YM}$ in the dual field theory. In this background let us introduce $N_f$ D$q$-probe branes that wrap ${t,x,u}$ and the remaining bulk space-time dimensions (along with $q-d$ of the compact directions) and assume that it has a trivial profile. 

We work with the following gauge potential
\begin{eqnarray}
A_x = - E t + a_x(u) \ , \quad A_t = a_t(u) \ .
\end{eqnarray}
The effective temperature and the effective chemical potential can be calculated, {\it e.g.}~using (\ref{TIR}) and (\ref{muIR2}), to get
\begin{eqnarray}
\left. T_{\rm eff} \right|_{\mu_{\rm eff} = 0} & = & \frac{\sqrt{(d-1)}}{\sqrt{2} \pi }E^{1/2} \ , \label{Teff} \\ 
\mu_{\rm eff} & \sim & d_0 \ E^{\frac{2-d}{2}} \ . \label{mueff}
\end{eqnarray}
For simplicity, we have calculated $T_{\rm eff}$ in the absence of any chemical potential and charge density. Thus $T_{\rm eff}$ is trivially fixed: it involves only one scale, the electric field. From (\ref{Teff}), we observe that $T_{\rm eff} \to \{0, \infty \}$ as $E \to \{0, \infty \}$, irrespective of the number of dimensions.

The scaling of the chemical potential can be determined by noting that, as a result of the equations of motion and hence a dynamical input is needed, $\mu_{\rm eff}$ must be proportional to $d_0$. Given this, the scaling with the electric field in (\ref{mueff}) follows immediately from dimensional analysis. Note that, for $d=2$, $\mu_{\rm eff}$ will be calculated as a special case. For $d > 2$, $\mu_{\rm eff} \to \{\infty, 0\}$, in the limit $E\to \{0, \infty\}$. In the $E \to \infty$ limit, the osm event horizon is located at the conformal boundary and it requires no work to introduce a charged particle in the system, resulting in $\mu_{\rm eff} \to 0$. On the other hand, when $E \to 0$, the osm horizon shrinks to zero-size opening up the entire AdS-space along which the work needs to be done and thus $\mu_{\rm eff} \to \infty$.

Let us now turn towards regularization of the free energy. In what follows, we will consider the case with $\mu_{\rm eff} = 0$, since the divergence structure is not modified by a non-vanishing chemical potential. The Euclidean on-shell action is:
\be
T_{\rm eff} \ S_{\rm DBI}^{({\rm E})} = N_f \tau_q \Omega_{q-d} \int_0^{u_*} du \  \frac{u_*^{d-3}}{u^{d+1}}\left(\frac{u_*^4 - u^4}{u_*^{2(d-1)} - u^{2(d-1)}}\right)^{1/2} \ ,
\ee
which diverges in the limit $u \to 0$. To regularize this divergence, let us introduce the usual counter-term:
\be
L_1 = \frac{N_f \tau_q}{d} \sqrt{-\det \gamma}\ ,
\ee
where, $\gamma_{\mu \nu}$ is the induced metric on the non-compact part of the $q$-brane at $u=\epsilon$. In what follows, we will introduce additional counter terms, as needed.

Subsequently, working in the Lorentzian picture, the renormalized stress tensor is:
\be
\langle T^\mu_\nu\rangle_{\rm ren} = \int_{u_{\rm min}}^{u_{\rm max}} du d^{q-m-1} \vec{c}\ \sqrt{-\det G}\ U^\mu_\nu
\ee
where, now $U^{\mu\nu}$ is given by
\be
U^{\mu\nu} = \frac{1}{\sqrt{- \det G}}\left(\frac{\delta S_{\rm total}}{\delta G_{\mu\nu}} + \frac{\delta S_{\rm total}}{\delta G_{\nu\mu}} \right)\ , \quad S_{\rm total} = S_{\rm DBI} + S_{\rm ct} \ ,
\ee
where $S_{\rm ct}$ represents all the counter terms. We will now discuss each dimension separately.

Let us now briefly review the proposed free energy. In view of (\ref{freeosm}) and the subsequent discussions, the free energy is:
\begin{eqnarray}
\cF & = & \left.  T_{\rm eff} \ \left[ S_{\rm DBI}^{({\rm E})} \right|_{\rm on-shell} + S_{\rm ct}^{(\rm E)} - \left. \left(N_f \tau_q \Omega_{q-d} \right) j a_x \right|_{u_{\rm max}} \right] \nonumber\\
& = & \left.  T_{\rm eff} \ \left[ S_{\rm DBI}^{({\rm E})} \right|_{\rm on-shell} + S_{\rm ct}^{(\rm E)} - \left(N_f \tau_q \Omega_{q-d} \right) \int_{u_{\rm min}}^{u_{\rm max}} du j a_x' \int d^{d-1}\vec{x} \int dt_{\rm E}\right] \ ,
\end{eqnarray}
where, $a_x(u_{\rm min}) = 0$, which is the case for $d \ge 3$. We will treat $d=2$ separately. Also note that
\begin{eqnarray}
&& S_{\rm DBI} = - N_f \tau_q \Omega_{q-d} \int dt du d^{d-1}\vec{x} \ \L\left( a_x', a_t', u\right) \\ 
&& S_{\rm DBI}^{({\rm E})} = N_f \tau_q \Omega_{q-d} \int dt_{\rm E} du d^{d-1}\vec{x} \ \L\left( a_x', a_t', u\right) \quad {\rm with} \quad t \to - i t_{\rm E} \ ,
\end{eqnarray}
and
\begin{eqnarray}
S_{\rm ct} = c_1 \int dt d^{d-1}\vec{x} \ L_{\rm ct} \ , \quad S_{\rm ct}^{(\rm E)} = - c_1 \int dt_{\rm E} d^{d-1}\vec{x} \ L_{\rm ct} \ .
\end{eqnarray}
Thus, in the simplifying case of $\mu_{\rm eff}=0$, we get
\be
\cF = \left( N_f \tau_q \Omega_{q-d} \right) \A(d) E^{d/2} \propto T_{\rm eff}^d \ ,
\ee
which also follows from dimensional analysis. To compute the number $\A(d)$, however, we need to consider specific examples. Interestingly, even in the presence of a non-vanishing chemical potential, the on-shell action is $d_0$-independent in general, upon using the relation between $j$ and $d_0$.

Let us now offer a couple of comments regarding the regularization-scheme. To do so, let us define the following notations: We denote a radial-slice of the closed-string metric by $\gamma_{\rm closed}$, and a similar radial slice of the open string metric by $\gamma_{\rm open}$. Since both the closed and the open string metric are asymptotically AdS, we can immediately write down the following candidate counter-terms:
\begin{eqnarray}
L_{\rm ct}^{\rm closed} = - \frac{N_f \tau_q \Omega_{q-d}}{d}\sqrt{- {\rm det} \gamma_{\rm closed}} \ , \quad {\rm or} \quad  L_{\rm ct}^{\rm open} = - \frac{N_f \tau_q \Omega_{q-d}}{d}\sqrt{- {\rm det} \gamma_{\rm open}} \ . \nonumber\\
\end{eqnarray}
In fact, both $\gamma_{\rm closed}$ and $\gamma_{\rm open}$ can be mixed to construct the counter-terms. Some examples of such can be written as:
\begin{eqnarray}
L_{\rm ct}(\eta) \sim \left(\frac{- {\rm det} \gamma_{\rm closed}}{- {\rm det} \gamma_{\rm open}}\right)^{\eta} \sqrt{- {\rm det} \gamma_{\rm open}} \ ,
\end{eqnarray}
which represents a one-parameter family of counter-terms, parametrized by $\eta$. Clearly, as far as regularization is concerned, any member of this infinite family would suffice. However, depending on the choice, the finite part of the regulated free energy will be different. Note that, in deriving the free energy and the components of the stress-tensor, a natural choice emerges: $\eta = 1/4$. It may so happen, because of an underlying symmetry of the DBI-action, this fixes the regularization ambiguity. At this point, we will merely point out the potential infinite degeneracy of the regularization scheme and leave it at that.

\subsection{The case of $d=3$}

This is a very simple and somewhat trivial case, but illustrative nonetheless. Here, the dual field theory is $(2+1)$-dimensional and we collectively denote the field theory coordinates as $\vec{x}$. Here we have
\begin{eqnarray}
S_{\rm DBI} & = & - \lim_{\epsilon\to 0} N_f \tau_q \Omega_{q-3} \int_{\epsilon}^{u_*} \frac{du}{u^4} \int d^3 \vec{x} = \lim_{\epsilon\to 0} \frac{N_f \tau_q \Omega_{q-3}}{3} \left(\frac{1}{u_*^3} - \frac{1}{\epsilon^3} \right)\int d^3 \vec{x} \ , \\
S_{\rm ct} & = & \lim_{\epsilon\to 0}  \frac{N_f \tau_q \Omega_{q-3}}{3} \int d^3\vec{x} \sqrt{- \det \gamma} = \lim_{\epsilon\to 0}  \frac{N_f \tau_q \Omega_{q-3}}{3} \frac{1}{\epsilon^3} \int d^3\vec{x} \ .
\end{eqnarray}
Thus
\begin{eqnarray}
S_{\rm total} = S_{\rm DBI} + S_{\rm ct} = \frac{N_f \tau_q \Omega_{q-3}}{3} \ E^{3/2} \int d^3\vec{x} \ .
\end{eqnarray}
where $u_*=E^{-1/2} = j^{-1/2}$. Thus, the renormalized stress tensor is given by
\begin{eqnarray}
\langle T^\mu_\nu \rangle_{\rm ren} = \langle T^\mu_\nu \rangle + \frac{N_f \tau_q \Omega_{q-3}}{3}\sqrt{\gamma}\delta^\mu_\nu \ .
\end{eqnarray}
This yields
\begin{eqnarray}
\langle T^0_0 \rangle_{\rm ren} = - \frac{2}{3} N_f \tau_q \Omega_{q-3} E^{3/2}\ , \quad \langle T^1_1 \rangle_{\rm ren} = \frac{N_f \tau_q \Omega_{q-3}E^{3/2}}{3 } = \langle T^2_2 \rangle_{\rm ren} \ .
\end{eqnarray}
Clearly, the field theory stress tensor is traceless. Finally, here we get: $\A(3) = - \frac{4}{3}$. It may be tempting to infer that, a lack of log-divergence in the free energy, somehow indicates that conformal invariance is still restored and this is further supported by the traceless-ness of the stress tensor. We will encounter further examples in $d=5, 6$ where the absence of a log-divergence indeed correlates with a traceless-ness of the stress tensor piece.

\subsection{The case of $d=4$}

Here, the dual field theory is $(3+1)$-dimensional. This case, particularly the one originating from considering a D$7$-probe in a D$3$-brane background has been analyzed in detail in \cite{Karch:2008uy}. As in the previous example, here we have
\begin{eqnarray}
S_{\rm DBI} & = & \lim_{\epsilon\to 0} N_f \tau_q \Omega_{q-4} \left( - \frac{1}{4\epsilon^4} - \frac{1}{2} E^2 \ln \left(\frac{\epsilon}{u_*}\right) + {\rm finite} \right) \int d^4 \vec{x} \ , \\
S_{\rm ct}^{(1)} & = & \lim_{\epsilon\to 0} \frac{N_f \tau_q \Omega_{q-4}}{4} \int d^4\vec{x} \sqrt{-\det \gamma} = \lim_{\epsilon\to 0} \frac{N_f \tau_q \Omega_{q-4}}{4 \epsilon^4} \int d^4\vec{x} \ , \\
S_{\rm ct}^{(2)} & = & - \lim_{\epsilon\to 0} \frac{N_f \tau_q \Omega_{q-4}}{4} \int d^4\vec{x} \sqrt{-\det \gamma} F_{\mu\nu}F^{\mu\nu}\ln \left(\frac{\epsilon}{u_0}\right) \ .
\end{eqnarray}
Here $u_0$ is an arbitrary scale that appears to make the ratio $\left( \epsilon/ u_0 \right)$ inside the logarithm a dimensionless one. Physically it appears from the logarithmic violation of conformal invariance. This yields
\begin{eqnarray}
\langle T^\mu_\nu \rangle_{\rm ren} = \langle T^\mu_\nu \rangle & + & \frac{N_f \tau_q \Omega_{q-4}}{4}\sqrt{- \det\gamma}\delta^\mu_\nu \nonumber\\
& + & N_f \tau_q \Omega_{q-4} \sqrt{- \det\gamma}\left(F^{\rho \mu}F_{\rho \nu} - \frac{1}{4}\delta^\mu_\nu  F_{\mu\nu}F^{\mu\nu}\right)\ln \left(\frac{\epsilon}{u_0}\right)\ .
\end{eqnarray}
Written explicitly, we get
\begin{eqnarray}
&& \langle T^0_0 \rangle_{\rm ren} = - N_f \tau_q \Omega_{q-4}  E^2 \left( a_0 + \frac{1}{2} \log\left( \frac{u_*}{u_0}\right) \right)  \ , \\
&& \langle T^1_1 \rangle_{\rm ren} =  N_f \tau_q \Omega_{q-4}  E^2 \left( a_1 - \frac{1}{2} \log\left( \frac{u_*}{u_0} \right) \right) \ ,\\
&& \langle T^2_2 \rangle_{\rm ren} = \langle T^3_3 \rangle_{\rm ren} =  N_f \tau_q \Omega_{q-4} E^2 \left( a_2 + \frac{1}{2} \log\left( \frac{u_*}{u_0} \right) \right) \ , \\
&& a_0=0.026\ , \qquad a_1=0.434\ , \qquad a_2=0.046 \ .
\end{eqnarray}
Clearly, in this case, the stress tensor is not traceless but the trace is independent of $u_0$. Finally, here we get: $\A(4) = - 0.507$.

Let us now offer a few comments on the coefficient of the log-divergence: the universal term in the free energy.\footnote{We will elaborate on this for the case of $d=2$ which we consider later.} When the background is kept at a non-vanishing temperature, $T$, the coefficient of the log-divergence --- which is simply proportional to $E^2$, and viewed purely as a function of $T$ and $T_{\rm eff}$ --- yields: 
\begin{eqnarray}
\cF_{\rm uni} \sim \left( T_{\rm eff}^4 - 3 T^4 \right) + T_{\rm eff}^2 \sqrt{T_{\rm eff}^4 + 3 T^4} \ .
\end{eqnarray}
First, setting $T=0$, we observe that $\cF_{\rm uni} \sim T_{\rm eff}^4$, which is essentially fixed on dimensional ground. In the limit $T/T_{\rm eff} \ll 1$, we get $\cF_{\rm uni} \sim \alpha T_{\rm eff}^4 - \beta T^4$; where $\alpha$ and $\beta$ are numerical constants. This form, while no longer follows from dimensional analysis, is particularly interesting, in that it appears to be a linear superposition of two disjoint Stefan-Boltzman law, each kept at a respective temperature. It is tempting to intuit that this universal piece therefore captures the ``heat flow" between the probe and the background sector. Thus identified, the generic heat flow is a non-linear function of both the temperatures. Finally, as a consistency check we observe that for $T_{\rm eff} = T$, we have $\cF_{\rm uni} = 0$.

\subsection{The case of $d=5$}

Now the exercise is repetitive and thus we only sketch it. There is no log-divergence, and the only divergence are order $1/\epsilon^5$ and $1/\epsilon$, which can be accounted for by counter-terms of the following form: $\sqrt{- \det \gamma}$ and $\sqrt{- \det \gamma} F_{\mu\nu} F^{\mu\nu}$, with appropriate coefficients. With this, the stress-tensor components are:
\begin{eqnarray}
&& \langle T^0_0 \rangle_{\rm ren} = b_0 E^{5/2} \ , \quad \langle T^1_1 \rangle_{\rm ren} = b_1 E^{5/2} \ , \\
&&  \langle T^2_2 \rangle_{\rm ren} = \langle T^3_3 \rangle_{\rm ren} = \langle T^4_4 \rangle_{\rm ren} = - b_2 E^{5/2} \ , 
\end{eqnarray}
where $b_0, b_1, b_2$ are positive numbers satisfying: $- b_0 - b_1 + 3 b_2 = 0$, which implies the stress tensor piece is traceless.

\subsection{The case of $d=6$}

It is interesting to explore this case, since, so far we have encountered the log-divergence only for even $d$. A straightforward calculation, however, yields that the divergences are only
\begin{eqnarray}
\O\left( \frac{1}{\epsilon^6}\right)  \quad {\rm and} \quad \O\left( \frac{1}{\epsilon^2}\right) \ ,
\end{eqnarray}
but no log-ones, even in the presence of non-vanishing electric field and background temperature. As in the case of $d=5$, these divergences are taken care of by the counter-terms: $\sqrt{- \det \gamma}$ and $\sqrt{- \det \gamma} F_{\mu\nu} F^{\mu\nu}$, with appropriate coefficients. Finally one gets
\begin{eqnarray}
&& \langle T^0_0 \rangle_{\rm ren} =  f_0 E^{3} \ , \quad \langle T^1_1 \rangle_{\rm ren} = f_1 E^{3} \ , \\
&& \langle T^2_2 \rangle_{\rm ren} = \langle T^3_3 \rangle_{\rm ren} = \langle T^4_4 \rangle_{\rm ren} = \langle T^5_5 \rangle_{\rm ren} = - f_2 E^{3} \ , 
\end{eqnarray}
where $f_0, f_1, f_2$ are positive numbers satisfying: $- f_0 - f_1 + 4 f_2 = 0$, which implies the stress tensor piece is traceless.

\section{AdS$_3$/CFT$_2$: a special case}

We will discuss the case of AdS$_3$/CFT$_2$, and specially the case of introducing a space-filling probe brane in the AdS$_3$-background. In this background, a gauge field may have special asymptotic behaviour, which may result in subtleties in identifying the gauge-gravity dictionary. These issues have been addressed before, see {\it e.g.}~\cite{Jensen:2010em}. In this section, building on the results of \cite{Jensen:2010em}, we will discuss the physics of the flavour degrees of freedom. The other pragmatic reason for treating this case separately is that most calculations can be analytically performed.

To that end, let us begin by writing down the metric:
\begin{eqnarray}
ds^2 = - \frac{1}{u^2}dt^2 + \frac{1}{u^2} dx^2 + \frac{1}{u^2}du^2 \ ,
\end{eqnarray}
where the CFT$_2$ lives in the $\{t, x\}$-plane. We will consider three cases: (i) $T\not = 0$, $E=0$; (ii) $T=0$, $E\not = 0$ and (iii) $T\not = 0$, $E\not = 0$. Here $T$ is the background temperature, {\it i.e.}~the one corresponding to the closed string metric.

In our subsequent discussions, we will turn of the chemical potential, to keep the discussion simple. However, let us discuss the implication of the open string metric regularity condition on the gauge field $a_t(u)$. First, by solving the equations of motion (\ref{eomat}) and (\ref{eomax}), we get
\begin{eqnarray}
a_t(u) & = & \alpha_t + d_0 \left[ \sqrt{1 +  u^2 E} + \log\left( \frac{u}{1 + \sqrt{1 + u^2 E}}\right)  \right] \ ,  \label{atd2} \\
a_x(u) & = & \alpha_x - \sqrt{d_0^2 + E} \left[ \sqrt{1 +  u^2 E} +\log\left( \frac{u}{1 + \sqrt{1 + u^2 E}}\right)  \right] \ . \label{axd2}
\end{eqnarray}
Here $\alpha_t$ and $\alpha_x$ are two integration constants that are related by the osm regularity condition: 
\begin{eqnarray}
\left. a_t + \Omega a_x \right|_{u_*} = 0 \quad \implies \quad \frac{\alpha_t}{\alpha_x} = -  \frac{d_0}{\sqrt{d_0^2 + E}}   \ .
\end{eqnarray}
Thus, clearly, if we set $\alpha_x = 0$ we immediately have $\alpha_t = 0$. To determine the chemical potential, we cannot immediately use the definition in (\ref{muIR2}), since the gauge field itself diverges logarithmically near the AdS-boundary. We can define the chemical potential to be the work done, subtracting off the log-divergence. In this scheme we obtain:
\begin{eqnarray}
\mu_{\rm eff} \sim d_0 \left( \sqrt{2} - 1 - \sinh^{-1}(1) - \frac{1}{2} \log (E) \right) \ . 
\end{eqnarray}
Clearly, it is straightforward to restore the factors of $\alpha'$ and AdS-radius $L$ in the argument of the log-term. Thus defined, $\mu_{\rm eff} \to \pm \infty$ as $E \to \{\infty, 0\}$. This is expected since it is hard to decouple the UV and the IR in low dimensions. Note that, while using the asymptotic ($u \to 0$) expansion of (\ref{atd2}) to compute the contribution coming from the conformal boundary of AdS, we identify the coefficient of the log-divergent term as the charge density\cite{Jensen:2010em}.

\subsection{$T\neq 0$, $E=0$}

In this case, the open string metric, the induced metric and the background closed string metric are all identical. This is a completely standard case, where the results are rather trivial. The bulk metric is given by
\begin{equation}
ds^2= -\frac{1}{u^2}\left(1-\frac{u^2}{u_{\rm H}^2}\right)dt^2+\frac{1}{u^2} dx^2+\frac{1}{u^2}\left(\frac{1}{1-\frac{u^2}{u_{\rm H}^2}}\right)du^2 \ ,
\end{equation}
that has the following temperature:
\begin{equation}
T=\frac{1}{2\pi u_{\rm H}} \ .
\end{equation}
The on-shell DBI action, corresponding to space-filling probe branes, is now given by
\be
S_{\rm DBI} = - N_f \tau_q \Omega_{q-2} \int_\epsilon^{u_{\rm H}} d^2 x \frac{du}{u^3} = - \frac{N_f \tau_q \Omega_{q-2}}{2}\left(\frac{1}{\epsilon^2}-\frac{1}{u_{\rm H}^2} \right)\int d^2 x\ .
\ee
Let us now renormalize the on-shell action by adding the counter-term
\be
S_1 = \frac{N_f \tau_q \Omega_{q-2}}{2} \int d^2 x \sqrt{\gamma} = \frac{N_f \tau_q \Omega_{q-2}}{2} \left(\frac{1}{\epsilon^2}-\frac{1}{2u_{\rm H}^2}\right)\int d^2 x\ .
\ee
Therefore,
\be
S_{\rm DBI} + S_1 = \frac{N_f \tau_q \Omega_{q-2}}{4 u_{\rm H}^2}\int d^2 x \ 
\ee
and hence the free energy is given by
\be
\cF = - \frac{N_f \tau_q \Omega_{q-2}}{4 \left(2 \pi\right)^{2}} T^2  \int dx \ .
\ee
The renormalized stress-tensor is
\begin{align}
\langle T^\mu_\nu \rangle_{\rm ren} = \langle T^\mu_\nu \rangle + \frac{N_f \tau_q \Omega_{q-2}}{2}\sqrt{\gamma}\delta^\mu_\nu\ ,
\end{align}
where
\be
\langle T^\mu_\nu \rangle = - N_f \tau_q \Omega_{q-2} \int_\epsilon^{u_{\rm H}}  \frac{du}{u^3} \delta^\mu_\nu = - \frac{N_f \tau_q \Omega_{q-2}}{2}\left(\frac{1}{\epsilon^2}-\frac{1}{u_{\rm H}^2} \right) \delta^\mu_\nu\ ,
\ee
and hence
\begin{align}
\langle T^\mu_\nu \rangle_{\rm ren} = \frac{N_f \tau_q \Omega_{q-2}}{4 u_{\rm H}^2}\delta^\mu_\nu\ .
\end{align}
%

\subsection{$T=0$, $E\neq 0$}

Let us now take $u_{\rm H} \rightarrow 0$, and excite an electric field on the space-filling probe brane. We will further set $a_t(u) =0$. The corresponding open string metric is given by
\begin{equation}\label{osmetric}
ds_{\rm osm}^2= -\frac{1}{u^2}\left(1-\frac{u^4}{u_*^4}\right)d\tau^2+\left(\frac{1}{u^2}+\frac{1}{u_*^2}\right) dx^2+\frac{1}{u^2}\left(\frac{1}{1-\frac{u^2}{u_*^2}}\right)du^2 \ ,
\end{equation}
where
\begin{equation}
E = \frac{1}{u_*^2} \ , \qquad  j=\frac{1}{u_*}\ , \qquad \sigma_0=u_* \qquad {\rm and} \qquad T_{\rm eff}=  \frac{E^{1/2}}{\sqrt{2}\pi}\ .
\end{equation}
Now, for the gauge mode $a_x(u)$, we can define a ``grand canonical" and a ``canonical" ensemble, as one would do for a non-vanishing $a_t(u)$. We discuss these cases below, following closely the analyses of \cite{Jensen:2010em}.

\subsubsection{Grand canonical ensemble}

The asymptotic behaviour of $a_x(u)$ is:
\begin{eqnarray}
&& a_x(u) = a_0 \ln \left(\frac{u}{u_0} \right) + a_1 + \ldots \ , \\
&& \implies \quad  \delta a_x(u) = \delta a_0 \ln (u/u_0) + \delta a_1 \ , \label{varax}
\end{eqnarray}
where, $u_0$ is some arbitrary scale. Under this variation, the DBI action changes in the following way
\be
\delta S_{\rm DBI} =  N_f \tau_q \Omega_{q-2} \int_{u=\epsilon} d^2 x \frac{\delta a_0 \ln (u/u_0) + \delta a_1}{u_*} \ ,
\ee
where we have ignored the boundary-term at the osm event horizon, since we eventually subtract that to have a well-defined variational problem. Let us now concentrate on the boundary variations $\delta a_0$ and $\delta a_1$. The divergence structure of $S_{\rm DBI}$ is given by
\begin{eqnarray}
S_{\rm DBI} = \lim_{\epsilon \to 0}\left[ - \frac{1}{2 \epsilon^2} +\frac{E}{2} \log\epsilon + {\rm finite}  \right] \left( N_f \tau_q \Omega_{q-2} \right) \int d^2 x \ .
\end{eqnarray}

To cancel the divergences, let us now consider the following counter-terms:
\begin{align}
&S_1 = \frac{N_f \tau_q \Omega_{q-2}}{2} \int_{u=\epsilon} d^2x \sqrt{\gamma} \ ,\\
&S_2 = \frac{N_f \tau_q \Omega_{q-2}}{2} \int_{u=\epsilon} d^2x \sqrt{\gamma} F_{u \nu}F^{u \nu}\ln \left(\frac{u_1}{\epsilon} \right)\ ,
\end{align}
where $u_1$ is some scale that we will fix later. Note that we could add a third counter-term proportional to $\int_{u=\epsilon} d^2x\sqrt{\gamma}F_{u \nu}F^{u \nu}$, which is finite. But since, $u_1$ is an arbitrary scale, this term can be absorbed in $S_2$. Also note that one could use the following counter-term
\be
\tilde{S}_2 = \frac{N \tau_q \Omega_{q-d}}{2 \ln \left(\epsilon/u_1\right)} \int d^2x \sqrt{\gamma}\gamma^{ij}A_iA_j|_{u=\epsilon} \ 
\end{equation}
instead of $S_2$.\footnote{This counter-term is not gauge invariant. However, as argued in \cite{Hung:2009qk}, in AdS/CFT the allowed gauge transformations take the form $\delta A = d\phi$, such that the asymptotic of $A$ remains the same. The leading term of $d\phi$ at the boundary  at best can be $\O (u)$. The gauge transformation of $\tilde{S}_2 \sim u$, which vanishes at the conformal boundary.} Now, under the variation (\ref{varax}), we get:
\begin{align}
\delta S_1=0 \ ,\qquad \delta S_2 = - N_f \tau_q \Omega_{q-2}\int_{u=\epsilon}d^2 x \frac{\delta a_0 \ln (\epsilon/u_1)}{u_*}\ .
\end{align}
Therefore, after ignoring the boundary term at $u=u_*$, we obtain,
\be
\delta S_{\rm ren} = \delta (S_{\rm DBI} + S_1 + S_2) =  \frac{N_f \tau_q \Omega_{q-2}}{u_*} \int_{u=\epsilon} d^2x\ \left( \delta a_0 \ln (u_1/u_0)+ \delta a_1\right) \ .
\ee
We want a theory that corresponds to the grand canonical ensemble of the dual field theory. We will work in the quantization, where, $a_1$ is the source and $a_0$ is the response. Therefore, we will fix $a_1$ on the boundary and $\delta S_{\rm ren}$ should vanish. That tells us, $u_1=u_0$. Also from the last equation, we can also figure out the one-point function of the current 
\be
\J \equiv \langle \J_x \rangle = \frac{\delta S_{\rm ren}}{\delta a_1} = \frac{N_f \tau_q \Omega_{q-2}}{u_*} = N_f \tau_q \Omega_{q-2} j \ .
\ee
The renormalized stress-tensor is:
\begin{align}
\langle T^\mu_\nu \rangle_{\rm ren} = \langle T^\mu_\nu \rangle + \frac{N_f \tau_q \Omega_{q-2}}{2} \sqrt{\gamma}\delta^\mu_\nu + N_f \tau_q \Omega_{q-2} \sqrt{\gamma}\left(-F^{u \mu}F_{u \nu}+\frac{1}{2}\delta^u_\nu  F_{u\nu}F^{\mu\nu}\right)\ln \left(\frac{u_0}{\epsilon} \right)\ . \label{tmunuren}
\end{align}
Explicitly,  we obtain:
\begin{align}
&\langle T^0_0 \rangle_{\rm ren} = - \frac{N_f \tau_q \Omega_{q-2} }{4} \left(\frac{2 \log (u_*/u_0)+2 \sqrt{2}-3+\log (4)-2 \sinh ^{-1}(1)}{u_*^2}\right)\ ,\label{st1}\\
&\langle T^1_1 \rangle_{\rm ren} = \frac{N_f \tau_q \Omega_{q-2} }{4} \left(\frac{2 \log (u_*/u_0) + 2 \sqrt{2} - 1 + \log (4) - 2 \sinh ^{-1}(1)}{u_*^2}\right)\ .\label{st2}
\end{align}
Before we proceed, let us make few comments about the scale $u_0$. At finite $\J$, there is a logarithmic violation of conformal invariance. If we perform a rescaling $u\rightarrow \lambda u$, a  term $\ln \lambda$ will appear in the action \cite{Bianchi:2001de}. This fact is reflected by the appearance of $\log u_0$ in the stress-tensor. However, this term drops out in the trace of stress-tensor
\be
\langle T^\mu_\mu\rangle_{\rm ren} = \frac{N_f \tau_q \Omega_{q-2} }{2 u_*^2} = \frac{1}{2}\J E^{1/2} \ .
\ee

Let us now compute the grand canonical free energy. To that end, we need to further add the following boundary term to cancel the non-zero contribution of the gauge field coming from the osm event horizon:
\begin{eqnarray}
S_3 & = & \int_{u=u_*} d^2 x \J a_x(u_*) \ , \quad {\rm hence} \\
S_{\rm ren} & = & S_{\rm DBI} + S_1 + S_2 + S_3 \ .
\end{eqnarray}
Thus the Gibbs' free energy is:
\begin{eqnarray}
\cF & = & \frac{N_f \tau_q \Omega_{q-2}}{4} \left(2\pi^2\right) T_{\rm eff}^2 \left( A + 2 \log \left(\sqrt{2} \pi u_0 T_{\rm eff} \right) \right) \int d x\ , \\
A & = & 5-6 \sqrt{2} + 2 \sinh^{-1}1 - \log 4 \ .
\end{eqnarray}
%

\subsubsection{Canonical ensemble}

To define the canonical ensemble, let us now add another boundary term to the above action:
\be
\bar{S}_{\rm ren} = S_{\rm DBI} + S_1 - S_2 + S_3 - \int_{u=\epsilon}\sqrt{\gamma}d^2x\ \gamma^{\mu \nu}a_{\mu}\J_\nu\ ,
\ee
where
\be
a = a_x(u) dx \ , \qquad \J = \frac{N_f \tau_q \Omega_{q-2}}{u_*} dx \ .
\ee
First, let us find out the on-shell $\bar{S}_{\rm ren}$
\be
\bar{S}_{\rm ren} = - \frac{N_f \tau_q \Omega_{q-2}}{4 u_*^2} \left(2\ln \left(\frac{u_*}{u_0} \right) + A \right) \int d^2 x \ .
\ee
Now the variations of $\bar{S}_{\rm ren}$ under (\ref{varax}) is:
\be
\delta \bar{S}_{\rm ren} = - N_f \tau_q \Omega_{q-2}\int d^2 x\ a_1  \delta a_0 =\int d^2 x\ a_1  \delta \J \ .
\ee
Therefore, the renormalized action $\bar{S}_{\rm ren}$ corresponds to the canonical ensemble, in which $a_0$ is the source, and the corresponding Helmholtz free-energy density is given by
\be
\cF_{\rm H} = \frac{N_f \tau_q \Omega_{q-2}}{4} \left( 2\pi^2 \right) T_{\rm eff}^2 \left(A -  2 \log \left(\sqrt{2} \pi u_0 T_{\rm eff}  \right) \right) \ .
\ee
%

\subsection{$T\neq 0$, $E\neq 0$}

In this case, we consider a space-filling probe brane in the AdS$_3$-BTZ background and we turn on a gauge field on the probe. The corresponding open string metric is given by
\begin{eqnarray} \label{osmbtz}
ds_{\rm osm}^2 = -\frac{1}{u^2}\left(1-\frac{u^2}{u_*^2}\right)\left(1+\frac{u^2}{u_*^2}-\frac{u^2}{u_{\rm H}^2}\right)d\tau^2 & + & \left(\frac{1}{u^2}+\frac{1}{u_*^2}-\frac{1}{u_{\rm H}^2}\right) dx^2 \nonumber\\
& + &\frac{1}{u^2}\left(\frac{1}{1-\frac{u^2}{u_*^2}}\right)du^2 \ .
\end{eqnarray}
Also
\begin{equation}
E = \frac{1}{u_*^2}\sqrt{1-\frac{u_*^2}{u_{\rm H}^2}} \ , \qquad  j = \frac{1}{u_*}\sqrt{1-\frac{u_*^2}{u_{\rm H}^2}} \ , \qquad \sigma_0=u_* \ .
\end{equation}
The closed string and the open string metric temperatures are given by
\begin{eqnarray}
&& T = \frac{1}{2\pi u_{\rm H}} \ , \qquad T_{\rm eff} = \frac{\sqrt{1-\frac{ u_*^2}{2u_{\rm H}^2}}}{\sqrt{2}\pi u_* }= \left\{ T^4+\frac{E^2}{4 \pi^4} \right\}^{1/4} \  \\
&& u_*=\frac{\sqrt{2}\pi}{E}\left(T_{\rm eff}^2-T^2 \right)^{1/2} \ .
\end{eqnarray}
%

\subsubsection{Stress tensor}

The renormalized stress tensor can be calculated using (\ref{tmunuren}), which yields:
\begin{align}
&\langle T^0_0 \rangle_{\rm ren} = - \frac{N_f \tau_q \Omega_{q-2} }{4 u_{\rm H}^2} \left[-\ln 4 + \frac{u_{\rm H}^2}{u_*^2}\left(1+\ln 4 -2\sqrt{2-\frac{u_*^2}{u_{\rm H}^2}} \right)-2\left(1-\frac{u_{\rm H}^2}{u_*^2} \right)\times \right.\nonumber\\
&\left. \left\{\ln \left(\frac{u_* u_{\rm H}}{u_0(u_{\rm H}+\sqrt{2u_{\rm H}^2-u_*^2})} \right)+\frac{u_{\rm H}}{u_*}\ln\left( \frac{u_{\rm H}^4+2 u_* \left(\sqrt{2 u_{\rm H}^2-u_*^2}+u_*\right) u_{\rm H}^2-u_*^4}{\left(u_{\rm H}-u_*\right) \left(u_{\rm H}+u_*\right){}^3}\right)\right\}\right],\\
&\langle T^1_1 \rangle_{\rm ren} = - \frac{N_f \tau_q \Omega_{q-2} }{4u_{\rm H}^2}\left[\ln 4 + \frac{u_{\rm H}^2}{u_*^2}\left(1-\ln 4 -2\sqrt{2-\frac{u_*^2}{u_{\rm H}^2}} \right) \right.\nonumber\\
&~~~~~~~~~~~~~~~~~~~~~~~~~~~~~\left. +2\left(1-\frac{u_{\rm H}^2}{u_*^2} \right)\left\{\ln \left(\frac{u_* u_{\rm H}}{u_0(u_{\rm H}+\sqrt{2u_{\rm H}^2-u_*^2})} \right)\right\}\right]\ .
\end{align}
The trace of the stress-tensor does not depend on the parameter $u_0$
\begin{align}
\langle T^\mu_\mu\rangle_{\rm ren} = & - \frac{N_f \tau_q \Omega_{q-2} }{2 u_*^2} \left[\left(\frac{u_{\rm H}^2-u_*^2}{u_* u_{\rm H}} \right)\ln\left( \frac{u_{\rm H}^4+2 u_* \left(\sqrt{2 u_{\rm H}^2-u_*^2}+u_*\right) u_{\rm H}^2-u_*^4}{\left(u_{\rm H}-u_*\right) \left(u_{\rm H}+u_*\right){}^3}\right)\right.\nonumber \\
&~~~~~~~~~~~~~~~~~~~~~~~~~~~~~~~~~~~~~~~~~~~~~~~~~~\left.+ \left(1 -2\sqrt{2-\frac{u_*^2}{u_{\rm H}^2}} \right)\right]\ .
\end{align}
It can be checked that in the limit $u_{\rm H}\rightarrow \infty$, we recover the $T=0$ results and in the limit $u_{\rm H}=u_*$, we get the $E=0$ results.

\subsubsection{Free energy}

Let us compute in the canonical ensemble. The renormalized action that is relevant for the canonical ensemble is the following
\be
\bar{S}_{\rm ren} = S_{\rm DBI} + S_1 - S_2 + S_3 - \int_{u=\epsilon}\sqrt{\gamma}d^2x\ \gamma^{\mu \nu}a_{\mu}\J_\nu \ ,
\ee
where, now $\J$ is
\be
\J = \frac{N_f \tau_q \Omega_{q-2}}{u_*}\sqrt{1-\frac{u_*^2}{u_{\rm H}^2}} dx \ .
\ee
Therefore, free energy density is given by
\begin{align}
\cF = & \frac{N_f \tau_q \Omega_{q-2}}{4} \left(-2j^2  \log \left(\frac{2u_* u_{\rm H}/u_0}{\sqrt{2 u_{\rm H}^2-u_*^2}+u_{\rm H}}\right)+\frac{1}{u_*^2}-\frac{2 \sqrt{2 u_{\rm H}^2-u_*^2}}{u_*^2 u_{\rm H}}\right.\nonumber \\
&~~~~~~~~~~~~~~~~~~~~~~~~~~~~~~~~~~~~~~~~~~\left.-\frac{4j^2u_{\rm H}}{u_*}\log\frac{(u_*+\sqrt{2u_{\rm H}^2-u_*^2})^2}{2\sqrt{u_{\rm H}^2-u_*^2} (u_{\rm H}+u_*)} \right) \ .
\end{align}
A few comments are in order: Note that, $\cF$ can be written as a function of $\{T, E\}$ or $\{T_{\rm eff}, E\}$. As noted earlier, at finite electric field, the logarithmic violation of conformal symmetry makes appearance in the $\left( \log u_0 \right)$-term. This makes the free energy a scheme-dependent quantity. However, the coefficient of the log-term is universal and is given by:
\be
\cF_{\rm uni} = \frac{N_f \tau_q \Omega_{q-2}j^2}{2} = N_f \tau_q \Omega_{q-2} \pi^2 \left(T_{\rm eff}^2-T^2 \right) \ . \label{Funi}
\ee
It is interesting to note that $\cF_{\rm uni} \ge 0$.

In \cite{BD2}, it has been derived that the in $(1+1)$-dim CFT, the heat transfer across the boundary of two sub-systems with temperatures $T_L$ and $T_R$, respectively, is universally given by
\begin{eqnarray}
J = \frac{c\pi}{12} \left( T_L^2 - T_R^2 \right) \ ,
\end{eqnarray}
where $c$ is the central charge of the $(1+1)$-dim CFT. As commented in \cite{BD2}, the above formula bears similarity to the Stefan-Boltzmann law for black body radiation, for each $T_L$ and $T_R$, which are then subtracted (similar to what was earlier observed for the case of $d=4$, in the limit $T/T_{\rm eff} \ll 1$). In view of this, (\ref{Funi}) seems to capture precisely this physics, if we further identify $c \propto N_f$. In other words, the universal coefficient of the log-term in the flavour sector free energy knows about the heat transfer between the closed string background geometry and the open string matter field. Even though we will not attempt to compute the back-reaction in this article, it is tempting to conjecture that the heat transfer in this system --- even though conformal symmetry is broken by the flavour configuration --- falls under the same universal description as in the conformal field theory.

It is incumbent upon us, to investigate further this universal piece {\it e.g.}~in the presence of a non-vanishing charge density. Having a non-vanishing background temperature renders the presentation somewhat unwieldy, so we set $T=0$. The universal coefficient is:
\begin{eqnarray}
\cF_{\rm uni} \propto \left( j^2 - d_0^2 \right) \propto E \propto \pi T_{\rm eff}^2 + T_{\rm eff} \sqrt{\pi^2 T_{\rm eff}^2 + 2 d_0^2} \ .
\end{eqnarray}
The physical origin of the expression above is not clear to us, other than making a plausibility-based claim that this is the heat-transfer in the presence of a charge density.

\section{Understanding the stress-tensor: a toy model}

Let us now review a simple, yet instructive, toy model that was studied in \cite{Karch:2008uy}. We review this to make direct connections with some of the results in \cite{Karch:2008uy}, which puts the discussion of the boundary stress-tensor part in right perspective. First we consider Einstein gravity with a negative cosmological constant in $(d+1)$-dimensions:
\be
S=\frac{1}{16\pi G_N}\int d^{d+1}x \sqrt{- g} \left( R+\frac{d(d-1)}{L^2} \right) \ .
\ee
We will study Schwarzschild-AdS$_{d+1}$ solution of the above action which has the form
\be\label{scads}
ds^2 = \frac{L^2}{u^2}\left(-\left(1-\frac{u^d}{u_{\rm H}^d} \right)dt^2 +\frac{du^2}{1-\frac{u^d}{u_{\rm H}^d}}+d\vec{x}_d^2\right) \ .
\ee 
The corresponding Hawking temperature and free energy density of the black hole solution is given by\cite{Karch:2008uy}
\be
T = \frac{d}{4\pi u_{\rm H}} \ , \quad f=-\frac{1}{16 \pi G_N}\left(\frac{4\pi}{d} \right)^d L^{d-1}T^d \ .
\ee

Any asymptotically AdS metric can be written in the Fefferman-Graham form\cite{fg}
\begin{equation}\label{feffermangraham}
ds^2={L^2 \over z^2}\left(g_{\mu\nu}(z,x)dx^{\mu}dx^{\nu}+dz^2\right) \ ,
\end{equation}
from which the dual CFT metric $ds^2=g_{\mu\nu}(x) dx^{\mu}dx^{\nu}$ can be directly read off as $g_{\mu\nu}(x)= g_{\mu\nu}(0,x)$. The full function $g_{\mu\nu}(z,x)$, specifically the sub-leading terms in an expansion in $z$, encodes data corresponding to the expectation value of the CFT stress-energy tensor $T_{\mu\nu}(x)$. Written explicitlyy, in terms of the near-boundary expansion
\begin{equation}\label{metricexpansion}
g_{\mu\nu}(z,x)=g_{\mu\nu}(x)+z^2 g^{(2)}_{\mu\nu}(x)+\ldots +z^d g^{(d)}_{\mu\nu}(x)+ z^d \log \left(z^2 \right) h^{(d)}_{\mu\nu}(x)+\ldots~.
\end{equation}
For the case in hand, the stress tensor for the $d$-dimensional boundary theory is given by \cite{dhss,skenderis,skenderis2}
\begin{equation}\label{graltmunu}
\left\langle T_{\mu\nu}(x)\right\rangle={ d\, L^{d-1} \over 16\pi G_{N}} g^{(d)}_{\mu\nu}(x) \ .
\end{equation}
For the metric (\ref{scads}), the stress tensor is obtained to be
\be
\langle T_{00}\rangle = \frac{d-1}{16 \pi G_N}\left(\frac{4\pi}{d} \right)^d L^{d-1}T^d \ , \qquad \langle T_{ij}\rangle= \frac{\delta_{ij}}{16 \pi G_N}\left(\frac{4\pi}{d} \right)^d L^{d-1}T^d \ .
\ee
Therefore, energy density $\rho$, pressure $p$ and entropy density $s$ are given by
\begin{align}
&\rho=\frac{d-1}{16 \pi G_N}\left(\frac{4\pi}{d} \right)^d L^{d-1}T^d \ ,\qquad p=\frac{1}{16 \pi G_N}\left(\frac{4\pi}{d} \right)^d L^{d-1}T^d\ , \\
&s=\frac{d}{16 \pi G_N}\left(\frac{4\pi}{d} \right)^d L^{d-1}T^{d-1}\ .
\end{align}
Let us now introduce a spacetime filling D-branes to this theory, and subsequently consider the back-reaction:
\be
S = \frac{1}{16\pi G_N}\int d^{d+1}x \sqrt{- g} \left( R+\frac{d(d-1)}{L^2}- \epsilon_f T_0 \right)\ ,
\ee
where $T_0$ is the tension of the brane and $\epsilon_f$ is a parameter which can be thought of as $N_f/N_c$.

\subsection{Exact results}

An exact solution of this theory can be obtained simply by replacing $L\rightarrow L'$ in (\ref{scads}), where,
\be
\frac{d(d-1)}{L'^2}=\frac{d(d-1)}{L^2}- \epsilon_f T_0 \ .
\ee
Therefore,
\be
L'=\frac{L}{\sqrt{1-\frac{\epsilon_f T_0 L^2}{d(d-1)}}} \ .
\ee
Thus that the temperature remains unchanged and all other quantities are computed simply by replacing $L\rightarrow L'$. However, our goal is to calculate only $\O(\epsilon_f)$ corrections, which are given by
\be
\Delta\langle T_{00}\rangle = \frac{d-1}{32 \pi G_N d}\left(\frac{4\pi}{d} \right)^d L^{d+1}T^d \left( T_0 \epsilon_f \right) \ , \quad \Delta\langle T_{ij}\rangle= \frac{\delta_{ij}}{32 \pi G_Nd}\left(\frac{4\pi}{d} \right)^d L^{d+1}T^d \left(T_0 \epsilon_f \right) \ . \label{deltaST}
\ee
The change in free energy is:
\begin{align}
&\Delta f=-\frac{1}{32 \pi G_Nd}\left(\frac{4\pi}{d} \right)^d L^{d+1}T^d \left( T_0 \epsilon_f \right) +\O(\epsilon_f^2)\ .\label{f}
\end{align}
%

\subsection{Results in the probe limit}

We will now consider the probe limit and compute the above quantities to compare with results obtained in the previous section. In the probe limit, let us first write down the DBI action
\be
S_{\rm DBI} = -\frac{ \epsilon_f T_0 }{16\pi G_N}\int d^{d+1}x \sqrt{- g} \ .
\ee
The on-shell action is divergent and hence we add a counter-term
\be
S_{\rm ct} = \frac{ \epsilon_f T_0 L}{16\pi G_N d}\int d^{d}x \sqrt{- \gamma} \ ,
\ee
where $\gamma_{ab}$ is the induced metric at $u=\epsilon$. Therefore the free-energy density, which is obtained from the Euclidean DBI, is 
\be
\Delta f_{\rm probe} = -\frac{1}{32 \pi G_Nd}\left(\frac{4\pi}{d} \right)^d L^{d+1}T^d \left(T_0 \epsilon_f \right) + \O(\epsilon_f^2)\ , \label{probeF}
\ee
which agrees with the $\O(\epsilon_f)$ term of the exact result (\ref{f}). Let us now compute the stress tensor in the probe limit. The renormalized stress tensor --- that is obtained from the renormalized probe action --- is given by
\be
\Delta \langle T_{ab}\rangle_{\rm ren}^{\rm probe} = \frac{\eta_{ab}}{32 \pi G_Nd}\left(\frac{4\pi}{d} \right)^d L^{d+1}T^d \left(T_0 \epsilon_f \right) +\O(\epsilon_f^2)\ . \label{stressP}
\ee

Comparing (\ref{f}) and (\ref{probeF}), we see that $\Delta f = \Delta f_{\rm probe}$; the free energy of the system is fully determined by the probe sector up to $\O(\epsilon_f)$. However, comparing (\ref{deltaST}) and (\ref{stressP}), it can be concluded that the total stress tensor contribution up to $\O(\epsilon_f)$ comes from both probe sector and the back-reacted geometry:
\be
\Delta T_{ab} = \Delta T_{ab}^{\rm back-reaction}+\Delta T_{ab}^{\rm probe}\ .
\ee
It is natural, as argued in \cite{Karch:2008uy}, to identify $\Delta T^{\rm probe}_{ab} =  \Delta T^{\rm flavour}_{ab}$. In this paper, we have studied the physics of $\Delta f_{\rm probe}$ and $\Delta T_{ab}^{\rm probe}$, which directly involve the open string metic data.

\acknowledgments

We would like to thank Roberto Emparan, David Mateos, Julian Sonner for useful conversations. AK would like to thank the warm hospitality of Max-Planck-Institut f\"{u}r Physik, M\"{u}nchen, Universit\'{e} de Gen\`{e}ve and Universitat de Barcelona for warm hospitality during various stages of this work. SK is supported by NSF grant PHY-1316222. We would also like to thank the people of India for their generous support in research basic sciences.

\appendix

\section{Open String Metric}

We have been concerned with the following rank-$2$ tensor
\begin{eqnarray}
\M = G + \left( 2 \pi \alpha' \right) e^{-\phi/2} F = G + \bar{F} \ . \label{defM}
\end{eqnarray}
Now, for a generic matrix $\M$,
\begin{eqnarray}
\delta \sqrt{- {\rm det} \M} \,  = \frac{1}{2} \sqrt{- {\rm det} \M} \, {\rm Tr} \left( \M^{-1} \delta \M \right) \ ,
\end{eqnarray}
In deriving the above, we need the following equalities
\begin{eqnarray}
\log \left( \det \M \right)  =  {\rm Tr} \log \M \ , \quad  \delta \left( {\rm Tr} \log \M \right) & = & {\rm Tr} \left( \M^{-1} \delta \M \right) \ .
\end{eqnarray}
Also, note that, for an invertible matrix $\M$, we have
\begin{eqnarray}
\delta \M^{-1} = - \M^{-1} \delta \M \M^{-1} \ . \label{MMinvrel}
\end{eqnarray}
Here, identifying the matrix $\M$ as in (\ref{defM}), we get
\begin{align}
&\M^{-1}  =  \S + \A \ , \label{Minv} \\
&\S^{ab}  =  \left( \frac{1}{G+\F} \right)_{\rm sym}^{ab} = \left( \frac{1}{G+\F} \cdot G \cdot \frac{1}{G-\F}\right)^{ab} \ , \\
&\A^{ab}  =  \left( \frac{1}{G+\F} \right)_{\rm anti-sym}^{ab} = - \left( \frac{1}{G+\F} \cdot \F \cdot \frac{1}{G-\F}\right)^{ab} \ , \label{Aab} \\
&\delta \M_{ab}  =  \delta G_{ab} + \delta \F_{ab} \ , \\
& \S_{ab} = G_{ab} - \left( \bar{F} \cdot G^{-1} \cdot \bar{F} \right)_{ab} \ .
\end{align}
One can also check that 
\be
\S^{ab}\S_{bc}=\delta^a_c\ , \qquad G_{ad}\S^{dc}G_{cb}\neq \S_{ab}\ .
\ee

Now, as defined in (\ref{flavourst}), we obtain
\be\label{backreaction}
\T^{MN} = - \left(\kappa_{10}^2 N_f \tau_q\right) e^{\gamma \phi} \frac{\sqrt{-\det\M}}{\sqrt{-\det G_{10}}} \S^{ab} \delta_a^M \delta_b^N \delta^{9-p}\left(\vec{c}-\vec{c}_0\right) \ ,
\ee
where, $\vec{c}=\vec{c}_0$ represents the location of $q$-branes. Also note that,
\be
\det (G+\F)\det(G-\F)=\det \S \det G
\ee
and one can easily check that $\det (G+\F)=\det(G-\F)$ and hence,
\be
-\det \M= \sqrt{\det \S \det G}\ .
\ee
So, we can write (\ref{backreaction}) as
\be
\T^{MN} = - \left(\kappa_{10}^2 N_f \tau_q\right) e^{\gamma \phi} \left( \frac{\sqrt{\det\S\det G}}{- \det G_{10}}\right)^{1/2}\S^{ab}\delta_a^M \delta_b^N \delta^{9-q}\left(\vec{c}-\vec{c}_0\right)\ ,
\ee
Note that the back-reaction of $q$-branes depends on the open-string metric $\S_{ab}$.



\end{document}